\documentclass[12pt,times]{nagauth}

\usepackage{ragged2e}

\usepackage{amssymb}
\usepackage{graphicx}
\usepackage{caption}
\usepackage{subfig}
\usepackage[utf8]{inputenc}
\usepackage{amsmath}
\usepackage{mathtools, cuted}
\usepackage{curves}
\usepackage[numbers]{natbib}
\usepackage[normalem]{ulem}
\usepackage{rotating}
\usepackage{color}
\usepackage[utopia]{mathdesign}
\usepackage[OMLmathrm,OMLmathbf,OMLmathsf,sfdefault=fav,scaled=0.875]{isomath}

\usepackage{setspace}
\newcommand\BibTeX{{\rmfamily B\kern-.05em \textsc{i\kern-.025em b}\kern-.08em
T\kern-.1667em\lower.7ex\hbox{E}\kern-.125emX}}

\usepackage[colorlinks,bookmarksopen,bookmarksnumbered,allcolors=black]{hyperref}

\def\volumeyear{2017}

\graphicspath{{figures/}}

\usepackage[textsize=tiny]{todonotes}
\usepackage{marginnote}
\makeatletter
\renewcommand\@biblabel[1]{#1.}
\makeatother

\newcommand{\mand}{\quad \mathrm{and} \quad}

\newcommand{\tr}{\mathrm{tr}\,}
\newcommand{\mathd}{\mathrm{d}}

\newcommand{\mbf}[1]{{\mathbf{#1}}}
\newcommand{\mbfs}{\boldsymbol}
\newcommand{\dcdot}{\mbf{\,:\,}}

\newcommand{\mrm}{\mathrm}

\newcommand{\uexp}[1]{$^{\mathrm{{#1}}}$}

\newcommand{\mvec}[1]{\mathsfbfit{#1}}
\newcommand{\mmat}[1]{\mathsfbfit{#1}}

\begin{document}


\runningheads{L.~Zhang, et al.}{A hyperviscoplastic model and undrained triaxial tests of peat}

\title{A finite-strain hyperviscoplastic model and undrained triaxial tests of peat}

\author{L.~Zhang\affil{1}, B. C. O'Kelly\affil{2} and T.~ Nagel\affil{3}$^,$\affil{4}\corrauth}

\address{\affilnum{1} Gavin and Doherty Geosolutions, Unit A2, Nutgrove Office Park, D14 X627, Ireland.\break\affilnum{2} Department of Civil, Structural and Environmental Engineering, Trinity College Dublin, College Green, Dublin 2, Ireland.\break\affilnum{3} Department of Environmental Informatics, Helmholtz Centre for Environmental Research GmbH - UFZ, Permoserstr. 15, 04318 Leipzig, Germany.\break\affilnum{4} Department of Mechanical and Manufacturing Engineering, Trinity College Dublin, College Green, Dublin 2, Ireland. }

\corraddr{Department of Environmental Informatics, Helmholtz Centre for Environmental Research GmbH - UFZ; thomas.nagel@ufz.de.}

\begin{abstract}
This paper presents a finite-strain hyperviscoplastic constitutive model within a thermodynamically consistent framework for peat which was categorised as a material with both rate-dependent and thermodynamic equilibrium hysteresis based on the data reported in the literature. The model was implemented numerically using implicit time integration and verified against analytical solutions under simplified conditions. Experimental studies on the undrained relaxation and loading-unloading-reloading behaviour of an undisturbed fibrous peat were carried out to define the thermodynamic equilibrium state during deviatoric loading as a prerequisite for further modelling, to fit particularly those model parameters related to solid matrix properties, and to validate the proposed model under undrained conditions. This validation performed by comparison to experimental results showed that the hyperviscoplastic model could simulate undrained triaxial compression tests carried out at five different strain rates with loading/unloading relaxation steps.
\end{abstract}


\maketitle

\section{Introduction}

Peat is considered a challenging natural material in geotechnical engineering practice. Its high water content, organic and fibrous nature cause difficulties in sampling and laboratory testing of peat materials. The large deformations \citep{Adams1961, FarrellEric2012, OKelly2013}, structural anisotropy \citep{Hendry2012}, time-dependent behaviour \citep{Edil1979} and great spatial variability in properties and fibrosity \citep{hobbs1986} observed in peat upon mechanical loading are indicative of the geomechanical complexities inherent in both experiments and numerical modelling. This study aims to find a rational ingress to the complex mechanical phenomena by setting up clear and plausible theories of the peat behaviour and propose a constitutive relationship for peat within the modern mechanics context adhering to principles established in the theory of materials.

The majority of experimental research on peat was focused on determining the undrained shear strength and the (long-term) time-dependent behaviour associated with drainage, i.e. consolidation, in one-dimensional (1D) settings. The undrained shear strength was analysed with Mohr-Coulomb friction-cohesion failure criteria. As peat structure and morphology are different from those of granular mineral soils, including the presence of fibres, the physical meanings of friction and cohesion are equivocal \citep{Boylan2008}. Corresponding to consolidation tests performed on peat, several models were developed or adopted from clay modelling, such as rheological models \citep{barden, Berry1972, Edil1982, Edil1984}, the $C_\alpha/C_c$ concept \citep{Mesri1977, Mesri1987,Mesri1997}, isotache models \citep{Haan1996,Vermeer1999,Leoni2008}, and elastic visco-plastic (EVP) models \citep{Yin1989,Yin1990,Yin1994,Yin1996,OLoughlin2001}. 

Early rheological models have reproduced some observed features which were proposed based on postulated peat consolidation mechanisms by considering the material's micro-structure, such as the two-level structural model for fibrous peat by Berry et al. \cite{Berry1972} and the rheological model proposed by Barden \cite{barden} which treated the volumetric and deviatoric stress response as a single process by employing a generalization in terms of the effective stress ratio. Due to the nonlinearity and mathematical complexity of the rheological models, curve-fitting techniques became a mainstream topic after the 1970s where two popular approaches prevailed. The $C_\alpha/C_c$ concept described a unique relationship between the index of secondary compression $C_\alpha$ and the compression index $C_c$ and led to a controversy on the uniqueness of the end of primary (EOP) consolidation as well as the so-called hypotheses A and B \citep{Fox1992,Fox1994,Mesri1987,Lefebvre1984,Graham1983,Degago2009,Degago2013}. Besides, the 1D consolidation/compression models developed based on either time-line theory \citep{Bjerrum1967} or isotache theory \citep{Suklje1957} were adopted to model peats.

EVP models for peat simulations were adopted from the constitutive models developed for clay \citep{Yin1989,Yin1994,Yin1996} based on the time-line concept and thus routinely adopted critical-state elastoplastic models for 1D small-strain conditions. An equivalent time was defined as the time needed to creep from a reference time line to the current state \citep{Yin1989} and a unique limit time line existed in strain--effective stress space at infinite equivalent time \citep{Yin1994}. By defining a reference time line for the current strain and effective stress, negative equivalent time could be obtained as a function of the strain--effective stress states, illustrating how heuristic many of these modelling approaches remained. The models developed based on the time-line concept \citep{Bjerrum1967,Garlanger1972,Janbu1969,Yin1994} directly took time as a variable in describing the rate-dependence. Such an approach causes difficulties in relating the time elapsed since loading to the intrinsically time-dependent or rather rate-dependent behaviour of a soil \citep{Haan1996}. In most cases, using time as an explicit variable driving material behaviour is not advisable in constitutive theory of materials, cf. Haupt \cite{haupt2000} and others. O'Loughlin \cite{OLoughlin2001} found that the EVP model overpredicted fibrous peat settlement. 

The isotache model \citep{Haan1996} proposed a unique relationship between effective stress, strain and strain rate by defining a linear relationship between Hencky strain and the natural logarithm of intrinsic time as well as a linear relationship between Hencky strain and the natural logarithm of effective stress. The Hencky strain measure was chosen also to account for the large strains experienced by peat. Time as a variable was replaced by strain rate in the isotache model, however, an intrinsic time as a material parameter needed to be defined. This type of models was constrained by its curve-fitting origin from tests at a constant level of stress in 1D conditions \citep{Liingaard2004}. Albeit successful in modelling peat oedometer tests, the isotache model has difficulties in simulating additional loading conditions other than the original loading-unloading fitted curves, such as unloading relaxation \citep{DenHaan2001}. 

General constitutive laws describing the stress--strain--time relationships under any possible loading conditions have been derived from mechanisms which could plausibly be inferred to cause the observed constitutive behaviour \citep{Liingaard2004}. To account for the rate-dependence, the overstress theory \citep{perzyna1966,Adachi1987} and the nonstationary flow surface theory \citep{Olszak1966} are two perspectives which split rate-dependent elastoplasticity into rate-independent elastic and rate-dependent plastic parts. Neither of the theories are able to describe the relaxation process initiated within the yield surface due to the inherent combination of the viscous with plastic effects. In recent years, several elastoviscoplastic models were proposed for peat, such as the 1D small strain elastoviscoplastic model by Madachi and Gajo \cite{Madaschi2015}, the 3D small strain hyperplastic model by Boumezerane et al. \cite{Boumezerane} and the small strain kinematic bubble model by Boumezerane \cite{Boumezerane2014}. The models for peat reported in the literature suffered from a lack of experimental evidence---particularly regarding plastic yield criteria for peat---and consequently inherited certain concepts from the conventional small-strain approaches used in clay modelling. 

This paper adopts work performed for hyperviscoelastic modelling of biological tissues \citep{Gorke2010,Nagel2012} and extends it by a plastic component, within a thermodynamically consistent framework for simulating finite-strain rate-dependent peat behaviour under deviatoric conditions. The rationale for starting from deviatoric conditions will be explained in Section~\ref{sec:methodology}. Deviating from conventional approaches to modelling plastic deformations of peat, a plastic flow rule without a yield criterion was adopted \citep{Haupt2001}, see Section~\ref{sec:constitutive}. A numerical implementation of the devised model was realized relying on features like the Kelvin mapping \citep{Nagel2016}, nested Newton-Raphson iterations as well as implicit time integration and is described in Section~\ref{sec:numerical}. The viscoelastic parameters were calibrated from undrained triaxial relaxation tests and the elatoplastic parameters were calibrated from undrained loading-unloading test strain irrecoverability, see Section~\ref{sec:fits}. An initial validation of the constitutive model was realized in undrained triaxial testing of an undisturbed peat at six strain rates with loading/unloading relaxations and is presented in Section~\ref{sec:validation}. The article closes with a discussion and perspectives on future work in Sections \ref{sec:disc} and \ref{sec:concl}.

\section{General Methodology}
\label{sec:methodology}

This study adopts salient features of the material theory outlined by Haupt \cite{haupt2000}. The total stress ($\mbf{S}$) will be additively decomposed into a rate-independent equilibrium (elastoplastic) part ($\mbf{S}_\mrm{eq}$) and a rate-dependent (viscous) overstress ($\mbf{S}_\mrm{ov}$) in Eq.~\eqref{eq:stress_split}. The term \textit{equilibrium} refers to the thermodynamic state of the rate-dependent material in which all dissipative processes have ceased provided constant external conditions. The equilibrium stress is assumed to be a rate-independent functional of the process history, whereas the overstress is a rate-dependent function of the process history. 

\begin{equation}
\mbf{S} = \mbf{S}_\mrm{eq} + \mbf{S}_\mrm{ov}
\label{eq:stress_split}
\end{equation}

The overstress $\mbf{S}_\mrm{ov}$ in Eq.~\eqref{eq:stress_split} then has to disappear asymptotically for sufficiently slow processes or relax to zero during the course of any static continuation \citep{haupt2000}. The equilibrium state test, therefore, can be achieved by termination relaxations or defined by tests carried out at sufficiently slow rates. This has been experimentally illustrated in undrained triaxial relaxation tests on clay \citep{Silvestri1988}. Additionally, Graham et al. \cite{Graham1983} found that the strain-rate influence on the disturbed shear strength of clays appeared to be independent of soil plasticity or stress history during laboratory consolidation, and the gradient of the shear stress--$\log$\,(strain rate) relationship decreased markedly at low strain rates with a threshold strain rate of 0.2\,\%/hour for Haney clay \citep{Vaid1977} and about 0.05\,\%/hour for Drammen clay \citep{Berre1973}. A similar concept was defined as static yield surface by Sheahan \cite{Sheahan1995}, where a minimum undrained strength was eventually reached regardless of further strain-rate reductions, i.e. an equilibrium state was reached after the termination of creep tests. 

Summarising the observed geomechanical behaviour reported in the literature, undrained peat can be categorised as a rate-dependent material with thermodynamic equilibrium hysteresis, corresponding to a viscoplastic constitutive model. A hyperviscoelastic model was adopted motivated by the success of similar material models in the description of elastomeric \citep{Lion1997a,Haupt2001} and biological materials \citep{Gorke2010,Nagel2012}. Particularly the latter share some fundamental defining features with peat, such as a highly hydrated solid matrix being composed of fibre-reinforced organic matter. The hyperviscoelastic model was extended by incorporating plasticity to better account for the observed undrained peat behaviour. The proposed model was calibrated and validated with undrained triaxial tests of an undisturbed Sphagnum peat.

Quantifying and modelling the rate-dependence of fluid-filled porous media in general and peat in particular is complicated by the superposition of several non-linear effects. For example, the apparent viscoelasticity of such materials is due to both flow-dependent features caused by the momentum interaction between the solid matrix and the mobile pore fluid, and flow in-dependent phenomena associated with dissipative mechanisms in the solid matrix itself. An independent and physically based quantification of both contributions is practically only possible if first the flow-dependent features are removed from the picture by performing isochoric tests (fluid flow is primarily caused by volume changes of the pore space in a hydro-mechanical setting). Based on the resulting quantification of solid matrix rate dependence, non-isochoric tests such as oedometer tests can then be performed and modelled with a biphasic theory based on the effective stress concept to account for flow-dependent phenomena.

Hence, unconsolidated undrained triaxial tests were utilized here to calibrate the initial constitutive model on the merits of (1) the controllable boundary conditions; (2) avoiding the heterogeneity induced by consolidation \citep{halrelaxation12} and the Mandel-Cryer effect \citep{Zwanenburg2005}; and (3) avoiding rate-dependence of undrained peat resulting from the relative motion of pore water through the macro- and micro-porous structure. In a second step not presented here, the total stress constitutive model can be straightforwardly extended to an effective stress model as indicated in the previous paragraph. The laboratory testing plan was set up to allow the definition of a suitable equilibrium state for the peat material in undrained triaxial tests, to calibrate the rate-dependent and rate-independent components of the proposed model, and to provide additional experimental data for model validation. 

Note, that by way of convention, the sign conventions of mechanics (stresses and strains positive in tension, pressures positive in compression) are used throughout the modelling section. In the experimental section, as well as in all graphs presenting modelling results, stresses and strains are plotted with positive values in compression (sign convention of soil mechanics).

\section{Constitutive Model}
\label{sec:constitutive}

A macroscopic phenomenological model is proposed which can suitably describe the observed rate-dependent behaviour of peat materials at an engineering scale. As such, the model does not account for microstructural phenomena explicitly. As the current focus is on the rate-dependence and plasticity of the porous structure itself (cf. Section~\ref{sec:methodology}), the pore pressure was not accounted for by an explicit hydro-mechanical coupling but was lumped into the material response by means of a low compressibility resulting in quasi-isochoric behaviour. Furthermore, the constitutive model was derived firmly based on continuum mechanical and material theoretical principles \citep{haupt2000} to ensure a priori the thermodynamical consistency of the model results under arbitrary three-dimensional finite deformations.

\subsection{Rheological model}
\begin{figure}[th!]
\center
\includegraphics[width = 0.6\textwidth]{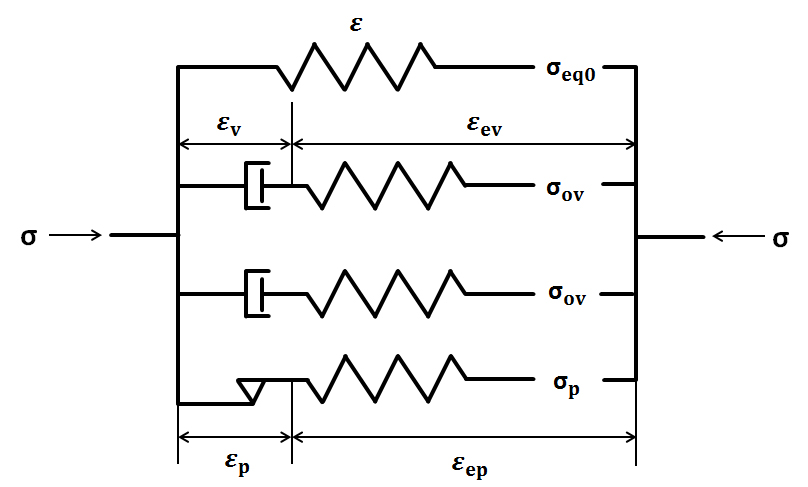}
\caption{Rheological network to visualise the hyperviscoplastic model structure}
\label{fig:modelstructure}
\end{figure}

The phenomenological rate-dependent constitutive model consisted of several parts, i.e. nonlinear elasticity, plasticity and rate-dependence. A parallel arrangement of a spring element, a spring-friction element and two Maxwell elements (Fig.~\ref{fig:modelstructure}) representing the elasticity, elasto-plasticity and visco-elasticity on two time scales, respectively, was chosen. This choice has successfully captured the complex behaviour of elastomeric \citep{Lion1997a,Haupt2001} and biological materials \citep{Gorke2010,Nagel2012}.

To facilitate an intuitive understanding of this structure, consider first a uniaxial setting under small-strain conditions. The total stress decomposes, according to Fig.~\ref{fig:modelstructure}, into elastic, elasto-plastic and visco-elastic components:

\begin{equation} \label{eq:additivestress}
\sigma = \underbrace{\sigma_\mrm{eq0} + \sigma_\mrm{p}}_\text{equilibrium state} + \sum_{i=1}^2 \sigma_\mrm{ov}^i
\end{equation}

For each layer of the parallel structure aside from the purely elastic spring, the total strain can be decomposed into elastic and inelastic parts:

\begin{equation} \label{eq:additivestrain}
\varepsilon = \varepsilon_\mrm{ep} + \varepsilon_\mrm{p} = \varepsilon_\mrm{ev}^i + \varepsilon_\mrm{v}^i
\end{equation}

\subsection{3D Formulation}

The three-dimensional finite-strain formulation of the constitutive model can be arrived at by a generalization of the uniaxial and small strain formulation \citep{Haupt1989,Iai2013}. In the sequel, material isotropy and isothermal conditions are assumed. An additive decomposition of the total Second Piola-Kirchhoff stress $\mbf{S}$ analogous to Eq.~\eqref{eq:additivestress} is introduced in accordance with the rheological model:

\begin{equation}
\mbf{S} = \mbf{S}_\mrm{eq} + \mbf{S}_\mrm{p} + \sum_{i = 1}^{2}\mbf{S}_\mrm{ov}^i
\end{equation}

For the finite-strain formulation, the additive decomposition of the total strain in Eq.~\eqref{eq:additivestrain} corresponds to a multiplicative decomposition \citep{Lee1969} of the deformation gradient, $\mbf{F}$, into elastic and inelastic parts:

\begin{equation}
\mbf{F} = \mbf{F}_\mrm{ep}\mbf{F}_\mrm{p} = \mbf{F}_\mrm{ev}^i \mbf{F}_\mrm{v}^i \label{eq:Fsplit}
\end{equation}

\noindent We further introduce the Green-Lagrange strain tensor, $\mbf{E}$, defined as

\begin{equation} \label{eq:GL}
\mbf{E} = \frac{1}{2} \left( \mbf{F}^\mrm{T} \mbf{F} - \mbf{I} \right) = \frac{1}{2}\left( \mbf{C} - \mbf{I} \right)
\end{equation}

\noindent with the right Cauchy-Green deformation tensor

\begin{equation} \label{eq:RCG}
\mbf{C} = \mbf{F}^\mrm{T} \mbf{F}
\end{equation}

\noindent Only in the intermediate configurations induced by the split in Eq.\,\eqref{eq:Fsplit}, the total strain decomposes additively into purely elastic parts of Green-Lagrangian type and purely inelastic parts of Almansi-Euler type, i.e.

\begin{align}
& \mbf{\epsilon} = \mbf{F}_\mrm{p}^\mrm{-T} \mbf{E} \mbf{F}_\mrm{p}^\mrm{-1} = \frac{1}{2} \left( \mbf{F}_\mrm{ep}^\mrm{T} \mbf{F}_\mrm{ep} - \mbf{F}_\mrm{p}^\mrm{-T} \mbf{F}_\mrm{p}^{-1} \right) = \mbf{\epsilon}_\mrm{ep} + \mbf{\epsilon}_\mrm{p}\\
& \mbf{\epsilon} = \mbf{F}_\mrm{v}^{i\mrm{-T}} \mbf{E} \mbf{F}_\mrm{v}^{i\mrm{-1}} = \frac{1}{2} \left( \mbf{F}_\mrm{ev}^{i\mrm{T}} \mbf{F}_\mrm{ev}^i - \mbf{F}_\mrm{v}^{i\mrm{-T}} \mbf{F}_\mrm{v}^{i{-1}} \right) = \mbf{\epsilon}_\mrm{ev}^i + \mbf{\epsilon}_\mrm{v}^i
\end{align}

\noindent with 

\begin{align}
\mbf{\epsilon}_\mrm{ep} = \frac{1}{2} \left( \mbf{F}_\mrm{ep}^\mrm{T} \mbf{F}_\mrm{ep} - \mbf{I} \right) \qquad &\text{and} \qquad \mbf{\epsilon}_\mrm{p} = \frac{1}{2} \left( \mbf{I} - \mbf{F}_\mrm{p}^\mrm{-T} \mbf{F}_\mrm{p}^{-1} \right)\\
\mbf{\epsilon}_\mrm{ev}^i = \frac{1}{2} \left( \mbf{F}_\mrm{ev}^{i\mrm{T}} \mbf{F}_\mrm{ev}^i - \mbf{I} \right) \qquad &\text{and} \qquad \mbf{\epsilon}_\mrm{v}^i = \frac{1}{2} \left( \mbf{I} - \mbf{F}_\mrm{v}^{i\mrm{-T}} \mbf{F}_\mrm{v}^{i{-1}} \right)
\end{align}

Objective strain rates in the current and intermediate configurations were obtained by Oldroyd rates via Lie-type time derivatives:

\begin{align}
&\overset{\Delta}{\mbf{\epsilon}} = \mbf{F}_\mrm{p}^\mrm{-T} \dot{\mbf{E}} \mbf{F}_\mrm{p}^{-1} = \dot{\mbf{\epsilon}} + \mbf{\epsilon}\mbf{l}_\mrm{p} + \mbf{l}_\mrm{p}^\mrm{T} \mbf{\epsilon}\\
&\overset{\Delta}{\mbf{\epsilon}} = \mbf{F}_\mrm{v}^{i\mrm{-T}} \dot{\mbf{E}} \mbf{F}_\mrm{v}^{i{-1}} = \dot{\mbf{\epsilon}}^i + \mbf{\epsilon}^i \mbf{l}_\mrm{v}^i + \mbf{l}_\mrm{v}^{i\mrm{T}} \mbf{\epsilon}^i
\end{align}

\noindent where the superimposed dot denotes the material time derivative, and the inelastic velocity gradients are given by

\begin{align}
\mbf{l}_\mrm{p} &= \dot{\mbf{F}}_\mrm{p} \mbf{F}_\mrm{p}^{-1} \mand 
\mbf{l}_\mrm{v}^i = \dot{\mbf{F}}_\mrm{v}^i \mbf{F}_\mrm{v}^{i-1}
\end{align}

\noindent In the intermediate configurations, the total rates $\overset{\Delta}{\mbf{\epsilon}}$ of the elastoplastic and viscoelastic components decompose additively into elastic and inelastic parts according to

\begin{align}
&\overset{\Delta}{\mbf{\epsilon}} = \overset{\Delta}{\mbf{\epsilon}}_\mrm{ep} + \overset{\Delta}{\mbf{\epsilon}}_\mrm{p} \label{eq:oldroyde}\\
&\overset{\Delta}{\mbf{\epsilon}}_\mrm{ep} = \dot{\mbf{\epsilon}}_\mrm{ep} + \mbf{\epsilon}_\mrm{ep}\mbf{l}_\mrm{p} + \mbf{l}_\mrm{p}^\mrm{T} \mbf{\epsilon}_\mrm{ep}
\\
&\overset{\Delta}{\mbf{\epsilon}}_\mrm{p} = \dot{\mbf{\epsilon}}_\mrm{p} + \mbf{\epsilon}_\mrm{p}\mbf{l}_\mrm{p} + \mbf{l}_\mrm{p}^\mrm{T} \mbf{\epsilon}_\mrm{p} = \frac{1}{2} \left( \mbf{l}_\mrm{p} + \mbf{l}_\mrm{p}^\mrm{T} \right)
\end{align} 

\noindent where $\overset{\Delta}{\mbf{\epsilon}}$ decomposes into $\overset{\Delta}{\mbf{\epsilon}}{}^i_\mrm{v}$ and $\overset{\Delta}{\mbf{\epsilon}}{}^i_\mrm{ev}$ analogously with the help of $\mbf{l}^i_\mrm{v}$.

The concept of dual variables \citep{Haupt1989} ensures that certain scalar physically relevant measures, such as the stress power, remain invariant under configurational transformations:

\begin{align}
&\mbf{S}_\mrm{p} \dcdot \dot{\mbf{E}} = \mbf{\tau}_\mrm{p} \dcdot \overset{\Delta}{\mbf{\epsilon}} = \mbf{\tau}_\mrm{p} \dcdot \left( \overset{\Delta}{\mbf{\epsilon}}_\mrm{ep} + \overset{\Delta}{\mbf{\epsilon}}_\mrm{p} \right)\\ 
&\mbf{S}^i_\mrm{ov} \dcdot \dot{\mbf{E}} ^i= \mbf{\tau}^i_\mrm{ov} \dcdot \overset{\Delta}{\mbf{\epsilon}} = \mbf{\tau}^i_\mrm{ov} \dcdot \left( \overset{\Delta}{\mbf{\epsilon}}{}^i_\mrm{ev} + \overset{\Delta}{\mbf{\epsilon}}{}^i_\mrm{v} \right) \label{eq:Sov:e}
\end{align}

\noindent where $\mbf{\tau}_\mrm{p}$ and $\mbf{\tau}^i_\mrm{ov}$ are the corresponding stresses in the intermediate configurations of the plastic and the $i^\mrm{th}$ viscoelastic second Piola-Kirchhoff stresses, i.e. 

\begin{align}
&\mbf{\tau}_\mrm{p} = \mbf{F}_\mrm{p} \mbf{S}_\mrm{p} \mbf{F}_\mrm{p}^\mrm{T} \label{eq:taup}\\
&\mbf{\tau}_\mrm{ov} = \mbf{F}^i_\mrm{v} \mbf{S}^i_\mrm{ov} \mbf{F}_\mrm{v}^{i\mrm{T}} \label{eq:tauov}
\end{align}
 
\noindent In the context of the rheological model shown in Fig.~\ref{fig:modelstructure}, the isothermal Helmholtz free energy is the sum of the strain energies of all spring elements. Thus, the specific Helmholtz free energy is assumed to decompose analogous to the stresses:

\begin{equation} \label{eq:freepsi}
\bar{\psi}\left( \mbf{E}, \mbf{\epsilon}_\mrm{ep}, \mbf{\epsilon}_\mrm{ev} \right) = \bar{\psi}_0 \left( \mbf{E} \right) + \bar{\psi}_\mrm{p} \left( \mbf{\epsilon}_\mrm{ep} \right) + \sum_{i=1}^2 \bar{\psi}^i_\mrm{ev}\left( \mbf{\epsilon}_\mrm{ev} \right)
\end{equation}

Thermodynamic consistency is ensured here by deriving the constitutive relations from the entropy inequality according to the Coleman-Noll procedure \citep{Coleman1963}. Demanding non-negative entropy production rates for an isothermal process, the Clausius-Duhem inequality reduces to the Clausius-Planck relation given in a Lagrangian setting by

\begin{equation} \label{eq:etpinequ}
\mbf{S} \dcdot \dot{\mbf{E}} - \rho_0 \dot{\bar{\psi}} \geq 0
\end{equation}

\noindent Insertion of the stress decomposition, Eq.~\eqref{eq:additivestress} along with the constitutive assumption \eqref{eq:freepsi}, yields

\begin{equation}
\begin{aligned}
& \left( \mbf{S}_\mrm{eq} -\rho_0 \frac{\partial \bar{\psi_0}}{\partial \mbf{E}} \right)\dcdot \dot{\mbf{E}} + \left( \mbf{\tau}_\mrm{p} - \rho_0 \frac{\partial \bar{\psi}_\mrm{p}}{\partial \mbf{\epsilon}_\mrm{ep}} \right) \dcdot \overset{\Delta}{\mbf{\epsilon}}_\mrm{ep} + \mbf{\tau}_\mrm{p} \dcdot \overset{\Delta}{\mbf{\epsilon}}_\mrm{p} +\\
& + \rho_0 \frac{\partial \bar{\psi}_\mrm{p}}{\partial \mbf{\epsilon}_\mrm{ep}} \dcdot \left( \mbf{\epsilon}_\mrm{ep} \mbf{l}_\mrm{p} + \mbf{l}^\mrm{T}_\mrm{p} \mbf{\epsilon}_\mrm{ep} \right) + \sum_{i = 1}^2 \left( \mbf{\tau}_\mrm{ov}^i - \rho_0 \frac{\partial \bar{\psi}^i_\mrm{v}}{\partial \mbf{\epsilon}^i_\mrm{ev}} \right) \dcdot \overset{\Delta}{\mbf{\epsilon}}^i_\mrm{ev} +\\
& + \sum_{i=1}^2 \left( \mbf{\tau}^i_\mrm{ov} \dcdot \overset{\Delta}{\mbf{\epsilon}}^i_\mrm{v} \right) + \sum_{n=1}^2 \left[ \rho_0 \frac{\partial \bar{\psi}^i_\mrm{v}}{\partial \mbf{\epsilon}^i_\mrm{ev}} \dcdot \left( \mbf{\epsilon}^i_\mrm{ev} \mbf{l}^i_\mrm{v} + \mbf{l}^{i\mrm{T}}_\mrm{v} \mbf{\epsilon}^i_\mrm{ev}\right) \right] \geq 0
\end{aligned}
\end{equation}

\noindent where use has been made of the concept of dual variables, cf. Haupt \cite{Haupt2001}, G{\"o}rke et al. \cite{Gorke2010} and others. Standard arguments lead to the stress relations:

\begin{align}
& \mbf{S}_\mrm{eq} = \rho_0 \frac{\partial \bar{\psi_0}}{\partial \mbf{E}} \label{eq:getSeq0}\\
& \mbf{\tau}_\mrm{p} = \rho_0 \frac{\partial \bar{\psi}_\mrm{p}}{\partial \mbf{\epsilon}_\mrm{ep}}\\
& \mbf{\tau}_\mrm{ov}^i = \rho_0 \frac{\partial \bar{\psi}^i_\mrm{v}}{\partial \mbf{\epsilon}^i_\mrm{ev}}
\end{align}

\noindent Assuming $\bar{\psi}_\mrm{p}$ and $\bar{\psi}^i_\mrm{v}$ are isotropic tensor functions of their arguments $\mbf{\epsilon}_\mrm{ep}$ and $\mbf{\epsilon}^i_\mrm{ev}$, respectively, i.e. $\psi(\mbf{A}) = \psi(\mbf{Q} \mbf{A} \mbf{Q}^\mrm{T})\ \ \forall \mbf{Q} \in \mrm{SO(3)}$, we find

\begin{align}
& \frac{\partial \bar{\psi}_\mrm{p}}{\partial \mbf{\epsilon}_\mrm{ep}} \dcdot \left( \mbf{\epsilon}_\mrm{ep} \mbf{l}_\mrm{p} + \mbf{l}^\mrm{T}_\mrm{p} \mbf{\epsilon}_\mrm{ep} \right) = 2 \mbf{\epsilon}_\mrm{ep} \frac{\partial \bar{\psi}_\mrm{p}}{\partial \mbf{\epsilon}_\mrm{ep}} \dcdot \overset{\Delta}{\mbf{\epsilon}}_\mrm{p}\\
& \frac{\partial \bar{\psi}^i_\mrm{v}}{\partial \mbf{\epsilon}^i_\mrm{ev}} \dcdot \left( \mbf{\epsilon}^i_\mrm{ev} \mbf{l}^i_\mrm{p} + \mbf{l}^{i\mrm{T}}_\mrm{v} \mbf{\epsilon}^i_\mrm{ev} \right) = 2 \mbf{\epsilon}^i_\mrm{ev} \frac{\partial \bar{\psi}^i_\mrm{v}}{\partial \mbf{\epsilon}^i_\mrm{ev}} \dcdot \overset{\Delta}{\mbf{\epsilon}}{}^i_\mrm{v}
\end{align}

\noindent leaving the residual dissipation inequality

\begin{equation} \label{eq:rmdinequality}
\rho_0 \left( \mbf{I} + 2 \mbf{\epsilon}_\mrm{ep} \right) \frac{\partial \bar{\psi}_\mrm{p}}{\partial \mbf{\epsilon}_\mrm{ep}} \dcdot \overset{\Delta}{\mbf{\epsilon}}_\mrm{p} + \sum_{i=1}^2 \rho_0 \left( \mbf{I} + 2 \mbf{\epsilon}^i_\mrm{ev} \right) \frac{\partial \bar{\psi}^i_\mrm{v}}{\partial \mbf{\epsilon}^i_\mrm{ev}} \dcdot \overset{\Delta}{\mbf{\epsilon}}{}^i_\mrm{v} \geq 0
\end{equation}

\noindent to be satisfied. Following Haupt \cite{Haupt2001}, non-negativity is ensured by specifying the plastic flow rule and the viscous strain rate in the corresponding intermediate configurations with non-negative parameters $c_\mrm{p}$ and $\eta^i_\mrm{v}$:

\begin{align}
& \overset{\Delta}{\mbf{\epsilon}}_\mrm{p} := c_\mrm{p} \rho_0 ||\overset{\Delta}{\mbf{\epsilon}}|| \left( \mbf{I} + 2 \mbf{\epsilon}_\mrm{ep} \right) \frac{\partial \bar{\psi}_\mrm{p}}{\partial \mbf{\epsilon}_\mrm{ep}} \label{eq:intfrp}\\
& \overset{\Delta}{\mbf{\epsilon}}{}^i_\mrm{v} := \frac{ \rho_0}{\eta^i_\mrm{v}} \left( \mbf{I} + 2 \mbf{\epsilon}^i_\mrm{ev} \right) \frac{\partial \bar{\psi}^i_\mrm{v}}{\partial \mbf{\epsilon}^i_\mrm{ev}}\label{eq:intfrov}
\end{align}

\noindent where $\hat{\mbf{C}}_\mrm{ep} = \mbf{I} + 2 \mbf{\epsilon}_\mrm{ep}$ and $\hat{\mbf{C}}^i_\mrm{ev} = \mbf{I} + 2 \mbf{\epsilon}^i_\mrm{ev}$ are the elastic deformation tensor in the corresponding intermediate configurations of right Cauchy-Green type. For the generalisation of the proposed constitutive relationships, the plastic and viscous flow rules are introduced as by proportionality between the inelastic strain rates and corresponding stresses. The information on the yielding of peat materials in undrained conditions in the literature is insufficient to motivate the choice of a particular yield surface and parameterise it sufficiently. To avoid parameterisation of yield surfaces and plastic potentials based on insufficient experimental data, and by reasonably assuming that peat exhibits plastic effects very early on during its loading response in undrained conditions, the present plasticity model requires neither a yield function nor a plastic potential and can be parameterized by one material parameter only. For simplification, the same Helmholtz free energy density functions, $\psi = \rho_0 \bar{\psi}$ for a modified Neo-Hookean model \citep{Gorke2010}, were adopted for all the spring elements in the constitutive model:

\begin{align}
&\psi_0 = \frac{C_1}{\alpha} \left[ e^{\alpha \left( I_1 - \ln I_3 - 3 \right)} \right] + D_2 \left( \ln I_3 \right)^2 \label{eq:psi}\\
&\psi_\mrm{p} = \frac{C_\mrm{1p}}{\alpha_\mrm{p}} \left[ e^{\alpha_\mrm{p} \left( I_1^\mrm{ep} - \ln I_3^\mrm{ep} - 3 \right)} \right] + D_\mrm{2p} \left( \ln I_3^\mrm{ep} \right)^2 \label{eq:psip}\\
&\psi^i_\mrm{v} = \frac{C^i_\mrm{1v}}{\alpha^i_\mrm{v}} \left[ e^{\alpha^i_\mrm{v} \left( I^{\mrm{ev}i}_{1} - \ln I^{\mrm{ev}i}_3 - 3 \right)} \right] + D^i_{2\mrm{v}} \left( \ln I^{\mrm{ev}i}_3 \right)^2 \label{eq:psiv}
\end{align}

\noindent where the material parameters $C_1$,$C_\mrm{1p}$ and $C^i_\mrm{1v}$ are elastic moduli affecting both the distortional and the volumetric behaviour; $D_2$, $D_\mrm{2p}$ and $D^i_\mrm{2v}$ quantify bulk behaviour of the material; and $\alpha$, $\alpha_\mrm{p}$ and $\alpha^i_\mrm{v}$ control the nonlinearity of the constitutive relationship; $I_1$, $I_3$ and $I_1^\mrm{ep}$, $I_3^\mrm{ep}$ and $I_1^{\mrm{ev}i}$, $I_3^{\mrm{ev}i}$ are the first and third principal invariants of the right Cauchy-Green type tensors $\mbf{C}$, $\hat{\mbf{C}}_\mrm{ep}$ and $\hat{\mbf{C}}^i_\mrm{ev}$, respectively. Expressing the invariants by quantities defined in the reference configuration yields \citep{Lion1997a,Gorke2010}:

\begin{align}
& I_1 (\mbf{C}) = \tr (\mbf{C}) = \mbf{C} \dcdot \mbf{I}; \qquad I_3 (\mbf{C}) = \det \mbf{C} = J^2\\
& I_1^\mrm{ep} (\hat{\mbf{C}}_\mrm{ep}) = \tr\left( \mbf{C} \mbf{C}^{-1}_\mrm{p} \right);  \qquad I_3^\mrm{ep} (\hat{\mbf{C}}_\mrm{ep}) = \det \left( \mbf{C}\mbf{C}^{-1}_\mrm{p} \right)\\
& I_{1}^{\mrm{ev}i} (\hat{\mbf{C}}^i_\mrm{ev}) = \tr \left( \mbf{C} \mbf{C}^{i{-1}}_\mrm{v} \right); \qquad I_{3}^{\mrm{ev}i} (\hat{\mbf{C}}^i_\mrm{ev}) = \det \left( \mbf{C} \mbf{C}^{i{-1}}_\mrm{v} \right)
\end{align}

With the Helmholtz free energy functions \eqref{eq:psip} and \eqref{eq:psiv}, and by pulling back the inelastic flow rules \eqref{eq:intfrp} and \eqref{eq:intfrov}, the flow rules in the reference configuration read:

\begin{align}
& \dot{\mbf{C}}_\mrm{p} = 2 c_\mrm{p} ||\dot{\mbf{C}}|| \left( \frac{\partial \psi_\mrm{p}}{\partial I_1^\mrm{ep}} \mbf{C} + I_3^\mrm{ep} \frac{\partial \psi_\mrm{p}}{\partial I_3^\mrm{ep}} \mbf{C}_\mrm{p} \right) \label{eq:frp}\\
& \dot{\mbf{C}^i_\mrm{v}} = \frac{4}{\eta_\mrm{v}^i}\left( \frac{\partial \psi^i_\mrm{v}}{\partial I_1^{\mrm{ev}i}} \mbf{C} + I_3^{\mrm{ev}i} \frac{\partial \psi_\mrm{v}^i}{\partial I_3^{\mrm{ev}i}} \mbf{C}_\mrm{v}^i\right) \label{eq:frv}
\end{align}

\noindent and the total second Piola-Kirchhoff stress becomes

\begin{equation}
\begin{aligned}
\mbf{S} &= \mbf{S}_\mrm{eq} + \mbf{S}_\mrm{p} + \sum_{i=1}^2 \mbf{S}_\mrm{ov}^i\\
		&= 2 \left(\frac{\partial \psi_0}{\partial I_1} \mbf{I} + I_3 \frac{\partial \psi_0}{\partial I_3} \mbf{C}^{-1}\right) +\\
		&+ 2 \left( \frac{\partial \psi_\mrm{p}}{\partial I_1^\mrm{ep}} \mbf{C}_\mrm{p}^{-1} + I_3^\mrm{ep} \frac{\partial \psi_\mrm{p}}{\partial I_3^\mrm{ep}} \mbf{C}^{-1} \right) + \\
		&+ 2 \sum_{i=1}^2 \left( \frac{\partial \psi^i_\mrm{v}}{\partial I_1^{\mrm{ev}i}} \mbf{C}_\mrm{v}^{i-1} + I_3^{\mrm{ev}i} \frac{\partial \psi^i_\mrm{v}}{\partial I_3^{\mrm{ev}i}} \mbf{C}^{-1} \right)
\end{aligned}
\end{equation}

%
%
%

\section{Numerical Implementation}
\label{sec:numerical}

To allow the simulation of laboratory tests under the assumption of homogenous stress and deformation states, the material model was implemented into a numerical algorithm solving for mechanical equilibrium \citep{Janda2017}. 
The proposed constitutive relations contain rate formulations which were discretised in time using an implicit backward Euler scheme. Kelvin mapping was employed to represent tensor coordinates by vectors (for second-order tensors) and matrices (for fourth-order tensors) while maintaining tensor character (as opposed to the Voigt mapping) \citep{Nagel2016}. Original tensors are denoted by bold symbols ($\mbf{E}$, $\mbfs{\sigma}$) while corresponding Kelvin-mapped quantities are designated by {\it sans-serif}\ italic font ($\mvec{E}$, $\mvec{\sigma}$).

The total Lagrangian formulation of the weak form and the global Newton-Raphson iteration scheme for non-linear material models in a full finite element context are elaborated in Bucher et al. \cite{Bucher2001}, Zienkiewicz et al. \cite{Zienkiewicz2000} and Nagel et al. \cite{Nagel2016,Nagel2017}. Here, due to the homogeneity assumption, a simplified version could be used.

\subsection{Global Newton and Local Stress Iterations}

Nonlinearities were resolved using two nested Newton iteration algorithms for the global and local levels, respectively. The global problem is given by balancing internal and external stresses $\mvec{S}_\mrm{int} = \mvec{S}_\mrm{ext}$, i.e. by decreasing the norm of the residual

\begin{equation}
	\mvec{r}_\mrm{g} = \mvec{S}_\mrm{int} - \mvec{S}_\mrm{ext}
\end{equation}
to below a tolerance: $\|\mvec{r}_\mrm{g}\| < \epsilon_\mrm{tol}$. Corresponding to the displacement in a finite element setting, the primary unknown of the global problem here is the Green-Lagrange strain tensor $\mvec{E}$. The resulting Newton-Raphson scheme to solve for $\mvec{r}_\mrm{g} = \mvec{0}$ introduces the global Jacobian (tangent) matrix

\begin{equation}
	- \mvec{r}_\mrm{g}^i = \mmat{C}^i\, \Delta \mvec{E}^{i+1}
	\label{eq:newton_c}
\end{equation}

The evolution equations \eqref{eq:frp} and \eqref{eq:frv} in the hyperviscoplastic model have to be integrated at each load step to calculate the increments of the inelastic internal tensor variables $\mbf{C}_\mrm{p}$ and $\mbf{C}_\mrm{v}^i$. The differential equations are discretised in time using the implicit Euler backward scheme formulated in \eqref{eq:backeuler} considering a general ordinary differential equation $\dot{y} = f(y)$ for a time step in the interval $[t, t+\Delta t]$:

\begin{equation} \label{eq:backeuler}
 y^{t+\Delta t} = y^t + \Delta t f^{t+\Delta t}
\end{equation}

\noindent The time discretisation of the differential and algebraic equations necessary to integrate the stress increment can be written in the form of Eq.~\eqref{eq:residual} by defining a residual vector $\mvec{r}$ and the state vector $\mvec{z} = [\mvec{S}^\mrm{T}, \mrm{internal\ variables}^\mrm{T}]^\mrm{T}$, where $\mvec{E}^i$ is the strain tensor at the $i^\mrm{th}$ global iteration and is considered as fixed during the local iterations

\begin{equation} \label{eq:residual}
\mvec{r}\left( \mvec{z},\mvec{E}^i \right) = \mvec{0}
\end{equation}

\noindent A Taylor series expansion of \eqref{eq:residual} considering only linear terms in the neighbourhood of a given iterative solution of $\mvec{r}_j$ and $\partial \mvec{r}/\partial \mvec{z}|_j$ yields:

\begin{equation}
 - \mvec{r}^j = \left.\frac{\partial \mvec{r}}{\partial \mvec{z}}\right|_j \Delta \mvec{z}^{j+1}
\end{equation}

\noindent Once the iteration converged, the use of the total differential of $\mvec{r}$ directly provides the consistent tangent matrix for the global iteration:

\begin{equation}
 \frac{\mrm{d} \mvec{r}}{\mrm{d} \mvec{E}^{t+\Delta t}} = \frac{\partial \mvec{r}}{\partial \mvec{E}^{t+\Delta t}} + \left(\left. \frac{\partial \mvec{r}}{\partial \mvec{z}}\right|_{t+\Delta t} \right) \frac{\mrm{d} \mvec{z}^{t+\Delta t}}{\mrm{d} \mvec{E}^{t+\Delta t}} = \mvec{0}
\end{equation}

\noindent In the resulting linear system, the first entry of the solution $\mrm{d}\mvec{z}/\mrm{d}\mvec{E}^{t+\Delta t}$:

\begin{equation}
\left(\left. \frac{\partial \mvec{r}}{\partial \mvec{z}}\right|_{t+\Delta t} \right)\frac{\mrm{d} \mvec{z}^{t+\Delta t}}{\mrm{d} \mvec{E}^{t+\Delta t}} = -\frac{\partial \mvec{r}}{\partial \mvec{E}^{t+\Delta t}}
\end{equation}

\noindent is the material contribution $\mathd \mvec{S}/ \mathd \mvec{E}$ to the tangent matrix $\mvec{C}^i$ required in Eq.~\eqref{eq:newton_c}. This approach of two nested Newton-Raphson iterations enables quadratic convergence of both the local and the global problem due to an algorithmically consistent linearisation \citep{simo2000}. 

For the proposed hyperviscoplastic model, the components of the state vector is  $\mvec{z} = [ \widetilde{\mvec{S}}^\mrm{T}, \mvec{C}_\mrm{p}^\mrm{T}, \mvec{C}^1_\mrm{v}{}^\mrm{T}, \mvec{C}^2_\mrm{v}{}^\mrm{T} ]^\mrm{T}$, where $\widetilde{\mvec{S}} = \mvec{S}/C_1$ is a dimensionless stress measure. The residual vector, Jacobian and the global tangent solution of the hyperplastic model is presented in Appendix\,\ref{sec:apdx}.

The numerical implementation of the hyperviscoplastic model was successfully verified in four steps against specific analytical solutions of the elastic, viscoelastic, and elastoplastic model components under suitable test conditions \citep{Zhang2017thesis}.

\section{Undrained Triaxial Tests}

Undrained triaxial tests of an undisturbed Sphagnum peat were carried out to calibrate and validate the proposed finite-strain constitutive model. The model was calibrated in two steps, viz. equilibrium state and rate-dependent components, respectively. Correspondingly, suitable undrained triaxial loading-unloading equilibrium test needed to be determined experimentally, and tests at different loading rates were carried out. The remainder of this section details how a suitable equilibrium state test for the peat material was defined and details the experimental plan for model calibration.

The undisturbed peat samples were collected from Clara bog, Ireland, from beneath the groundwater table (sampling method and detailed properties, cf. Zhang and O'Kelly \cite{Zhang2014a}). Some selective properties of the tested peat material are presented in Table\,\ref{tab:peatproperty}.

\begin{table}[th!]
\centering
\caption{Selective properties of the tested peat material \protect\cite{Zhang2014a}.} \label{tab:peatproperty}
\small
\begin{tabular}{p{3.2cm} p{2.5cm} p{1.5cm} p{2.5cm} p{1.5cm}}
\hline
Natural water content per dry mass [\%] & Specific gravity of solid constituents & Loss on ignition [\%] & Fibre content [\%] & pH\\
\hline
590 & 1.42 & 98.6 & 63.5 & 3.8\\
\hline
\end{tabular}
\end{table}

A standard triaxial test set-up that included two pressure-volume controllers (GDS Instruments Ltd.) was used for undrained loading-unloading and relaxation tests on vertical undisturbed peat specimens ($\emptyset = 38$\,mm, $H = 76$\,mm). Saturation was confirmed by a Skempton $B$ value in excess of 0.95 for all specimens. The effect of cell pressure was investigated by undrained loading-unloading tests up to 5\,\% axial strain with incremental cell pressures of 0\,kPa, 20\,kPa, 40\,kPa and 120\,kPa on an undisturbed peat specimen at an axial strain rate of 16.0\,\%/hour. The test results demonstrated that the cell pressure had a negligible effect on the strain recovery as well as on the differential stress--axial strain relationships of the loading-unloading tests in this strain region. Thus, a confining pressure of 0\,kPa was taken for all subsequent undrained triaxial tests.

\subsection{Undrained triaxial relaxation tests}

The transient behaviour of a soil is believed to be caused by viscous effects in the soil structure, resulting in a rate-dependent material response with observable features such as creep and stress relaxation. Since the viscous effects observed in soil deformation result from different material effects caused by intrinsic properties or phase interactions, the term "rate-dependent" provides a more accurate description than "time-dependent" and allows a wider perspective for interpretation. The rate-dependence of the tested undisturbed peat was studied in three steps for calibrating the proposed model: (1) two stress relaxation tests were carried out at two strain rates with pore water pressure measurement under undrained conditions; (2) comparison of a slower loading-unloading test to a loading-unloading test with step relaxations at a higher strain rate; (3) definition of equilibrium state test. 

\begin{figure}[th!]
\center
\includegraphics[scale=0.7]{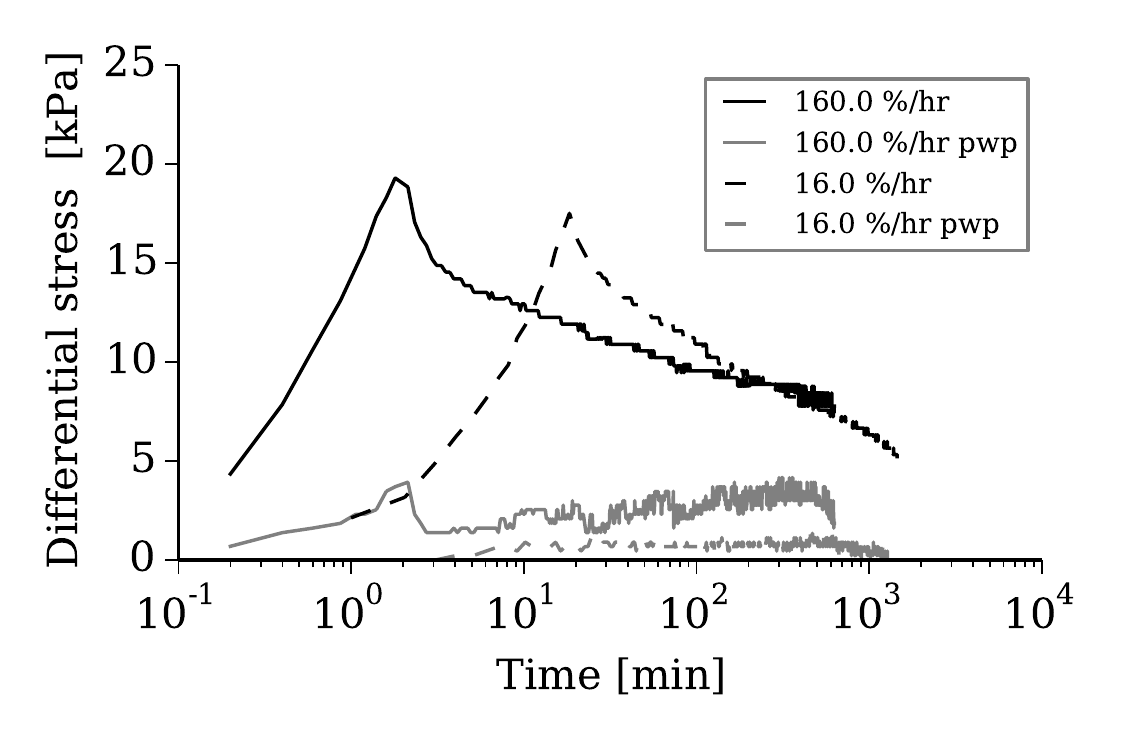}
\caption{Undrained relaxation tests on two undisturbed peat specimens at two strain rates. Black curves show differential stress, grey curves pore water pressure (pwp).}
\label{fig:2ratepwp}
\end{figure}

A triaxial relaxation test was carried out by stopping the displacement loading at a specific axial strain for a specimen tested at a given strain rate. Comparing to creep tests, which are commonly used for investigating material rate-dependence, relaxation tests record the stress change over time at a constant applied strain. Fig.~\ref{fig:2ratepwp} presents two triaxial compression tests to an axial strain of 5\,\% at axial strain rates of 160.0\,\%/hour and 16.0\,\%/hour followed by a 10 hour and a 24 hour relaxation phase, respectively, with pore water pressure measurement and zero cell pressure. The pore pressure variations were very small during the stress relaxations. A linear relationship can be found between the differential stress and the logarithm of time after an initial stress drop. Note that the "differential stress" is occasionally referred to as "deviator stress" in the literature. The latter terminology will be avoided here as the term "deviator" will be reserved for the actual deviatoric part of a tensor according to its mathematical definition which does not coincide with the differential stress.

\begin{figure}[th!]
\center
\includegraphics[scale=0.7]{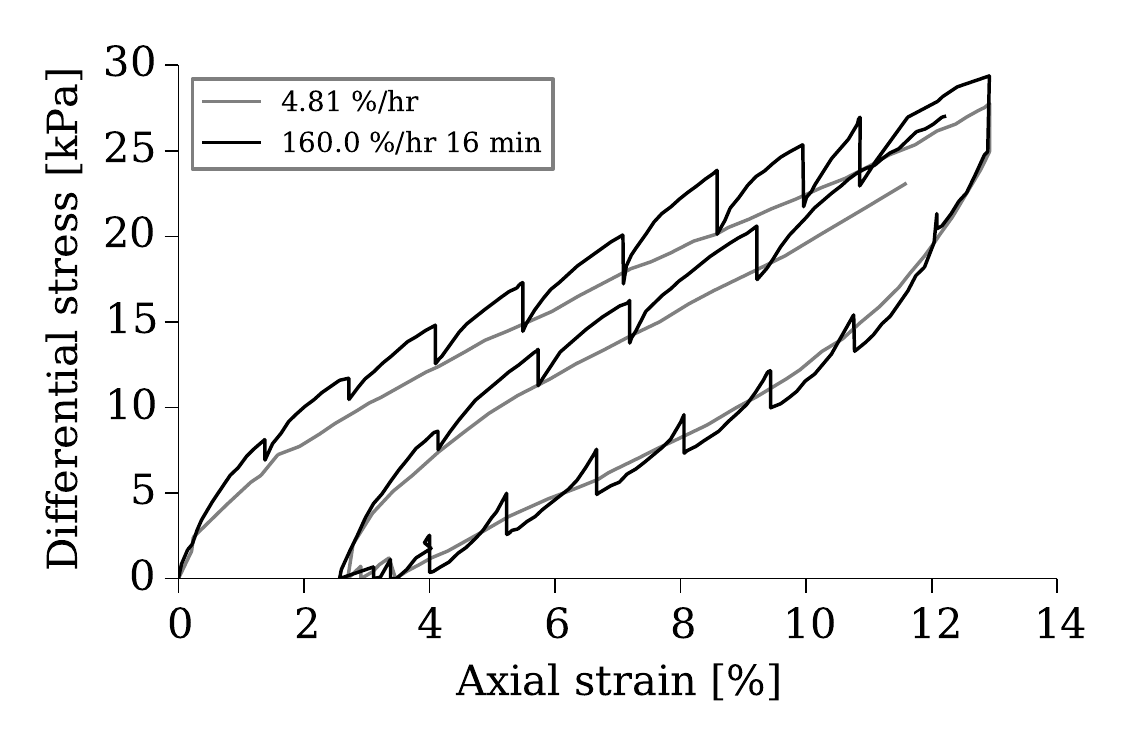}
\caption{Comparison of the monotonic 4.81\,\%/hour test with the 160.0\,\%/hour test with step relaxations.}
\label{fig:perfecttest}
\end{figure}

It is reasonable to postulate the existence of relaxation termination points for the undrained relaxation tests, i.e. an undrained equilibrium state, as the stress dissipation rate reduced exponentially with relaxation duration. An undrained triaxial loading-unloading test setup was used to compare the results of a quicker test with step relaxation periods to a slower but monotonic test under otherwise identical testing conditions. Specifically, the undrained loading-unloading triaxial test performed at 160.0\,\%/hour with step relaxations for 16\,min was compared with an undrained loading-unloading triaxial test at 4.81\,\%/hour. Fig.~\ref{fig:perfecttest} illustrates that the stress--strain curve obtained by the slow monotonic test was in close vicinity of the connected stress-relaxation end points of the fast step-relaxation (160\,\%\/hour, 16\,min) test. It was concluded that stress relaxation points of a higher-rate test with relaxations of a sufficient period of time can be approximated by a lower-rate test carried out at otherwise identical conditions. This observation was used for defining the equilibrium state.

\subsection{Equilibrium state}

\begin{figure}[th!]
\center
\includegraphics[scale=0.7]{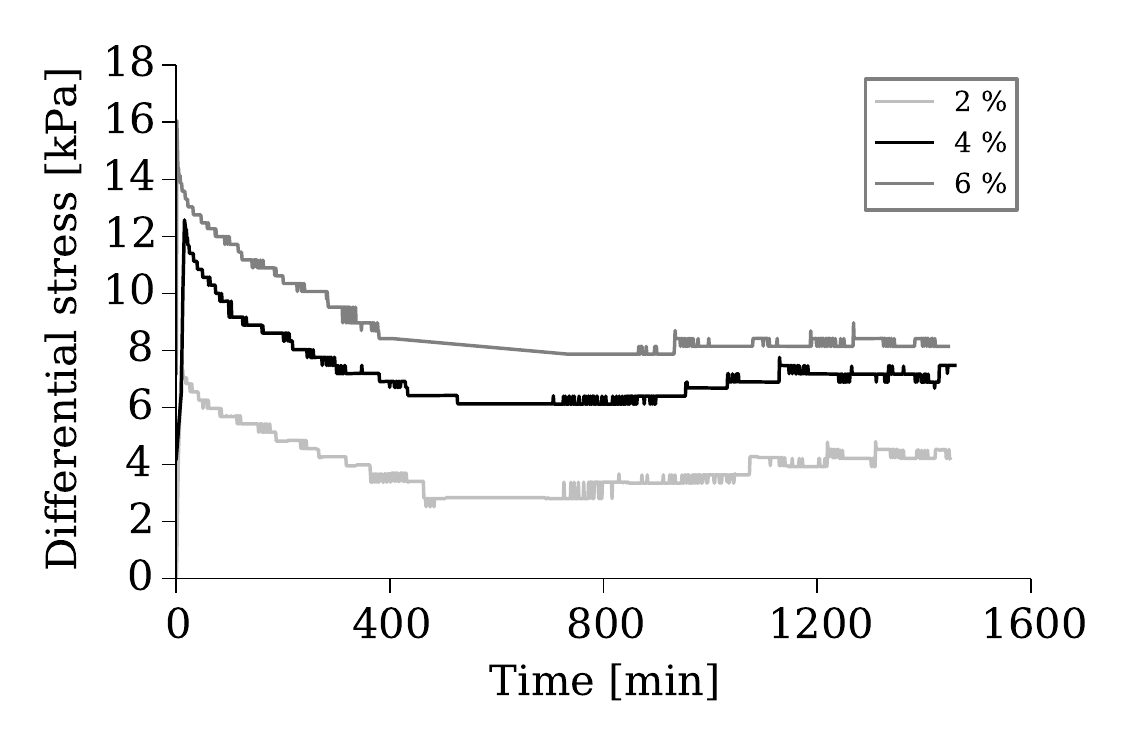}
\caption{Differential stress relaxation of undrained triaxial compression tests at axial strain levels of 2\,\%, 4\,\% and 6\,\%.}
\label{fig:steprelax}
\end{figure}

The equilibrium state can be obtained by connecting the stress relaxation termination points or approximated by tests carried out at a low strain rate. To determine the termination of stress relaxation, an undrained stress relaxation test at an axial strain rate of 16.0\,\%/hour was relaxed at strain levels of 2\,\%, 4\,\% and 6\,\% for 24 hours, respectively. The three stress relaxation phases are plotted against time on normal scale axis in Fig.~\ref{fig:steprelax}. The temperature influence on the stress relaxations was eminent by daily fluctuations. The temperature impact on the stress measurement turned out to be quite significant  with an average sensitivity of 2.11\,kPa/K making it difficult to precisely determine the relaxation termination points based on the current experimental equipment.

\begin{figure}[th!]
\center
\includegraphics[scale=0.7]{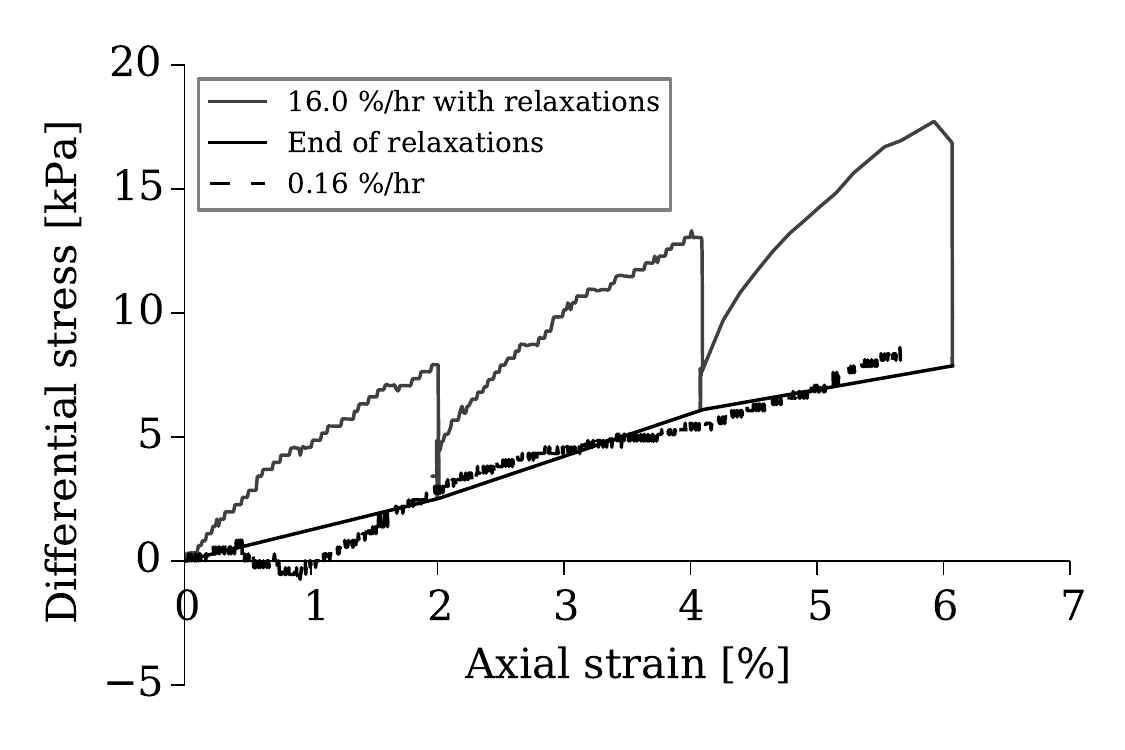}
\caption{Comparison of the monotonic 0.16\,\%/hour test with the 16.0\,\%/hour test with step-relaxation tests of 16~min duration.}
\label{fig:2equilibriums}
\end{figure}

Connecting the final stress relaxation points at each strain level and comparing the resulting curve to the stress--strain curve obtained by a monotonic undrained compression test at an axial strain rate of 0.16\,\%/hour, again yielded a close proximity between the two curves as observed in the previous paragraph for different experimental settings (Fig.~\ref{fig:2equilibriums}). Based on the finding that a lower-rate test can be used to approximate the relaxation termination points of a higher-rate test, we defined the undrained triaxial test at an axial strain rate of 0.16\,\%/hour as the ``equilibrium test'' for the undisturbed peat under undrained conditions for our current purposes.

\begin{figure}[th!]
\center
\includegraphics[scale=0.7]{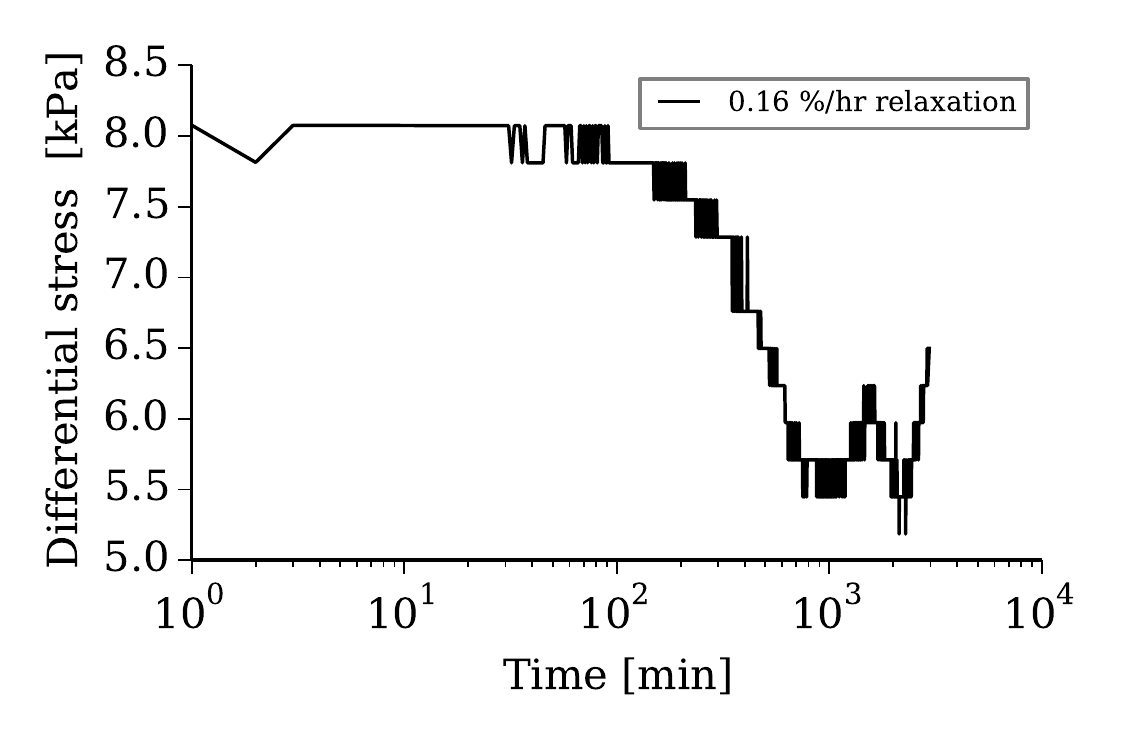}
\caption{Differential stress relaxation following compression to 6\,\% axial strain at 0.16\,\%/hour.}
\label{fig:equilitestrelax}
\end{figure}

By way of additional validation of this choice, a stress relaxation test was carried out at an axial strain rate of 0.16\,\%/hour to an axial strain of 6\,\%. Fig.~\ref{fig:equilitestrelax} presents the stress relaxation against logarithmic time. The average peak differential stress dissipation rate (slope gradient of the steepest stress drop in Fig.~\ref{fig:equilitestrelax}) was determined to be 0.0038\,kPa/min. Comparing this with the temperature sensitivity of the differential stress measured earlier, the highest amount of stress relaxation during a nine hour period was of the same magnitude as the stress variation instigated by a temperature change of 1.0\,K. Therefore, the definition of the equilibrium state of the undisturbed peat specimens in undrained triaxial tests at an axial strain rate of 0.16\,\%/hour can be reasonably justified by both the relaxation termination curve as well as the stress dissipation rate during relaxation. No data has been reported in the literature regarding the non-stationary yield surface or equilibrium state testing for peat in either drained or undrained conditions. However, Vaid et al. \cite{Vaid1977} reported a threshold strain rate of rate-independence for undrained Haney Clay of 0.2\,\%/hour, which is in close proximity to the value chosen here. 

\begin{figure}[th!]
\center
\includegraphics[scale=0.6]{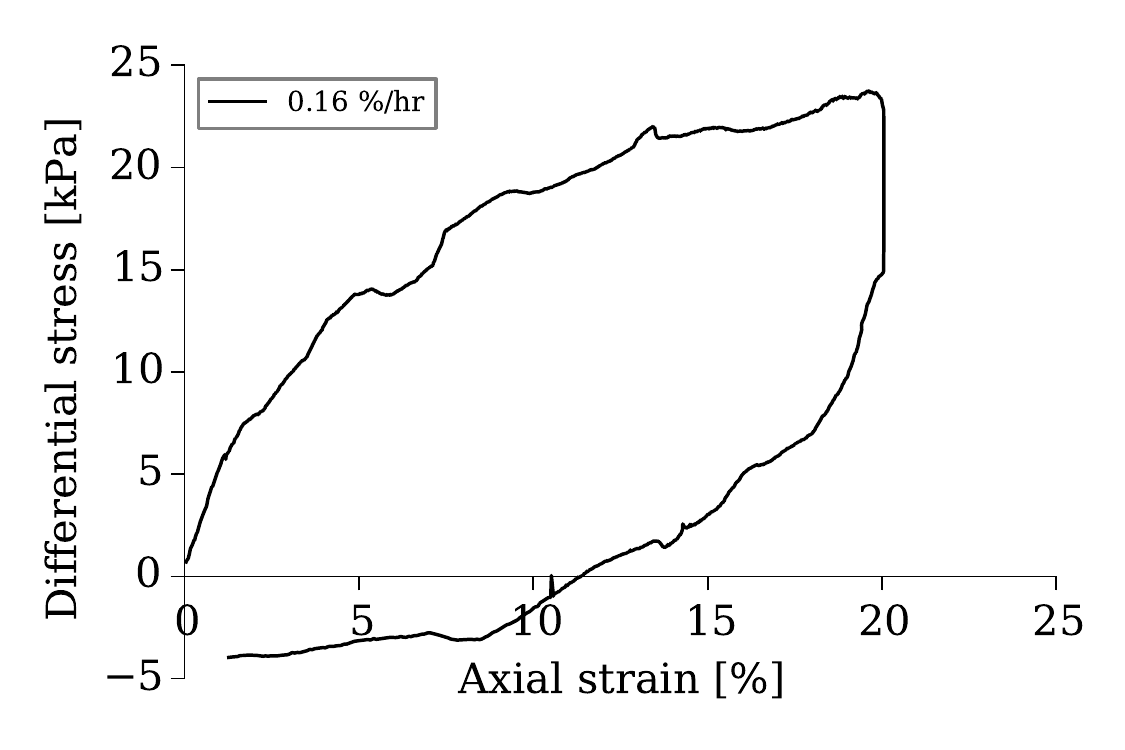}
\caption{Differential stress--axial strain relationship during an undrained triaxial test at 0.16\,\%/hour.}
\label{fig:equilu}
\end{figure}

Fig.~\ref{fig:equilu} presents the differential stress -- axial strain behaviour of the undisturbed peat at 0.16\,\%/hour. For a total axial strain of 20\,\%, the recovered strain based on the shape of the unloading curve was about 11.5\,\%. Fig.~\ref{fig:equilu} indicates strain irrecoverability in the defined equilibrium state test after a full strain cycle (equilibrium hysteresis) of the undisturbed peat material under undrained testing conditions. The difference between the differential stress values from Fig.~\ref{fig:equilu} and Fig.~\ref{fig:2equilibriums} was attributed to specimen variations as well as the temperature influence. Negative differential stress was recorded at the end of the unloading phase potentially resulting from the expulsion of the triaxial cell water (including the changes of hydrostaic pressure on the specimen and buoyancy on the load measurement) as well as inaccuracies due to temperature fluctuations. 

Finally note, that the testing conditions characterizing the equilibrium state may be redefined, given more accurate experimental data. While this may improve the model fits quantitatively, no fundamental differences to the subsequent results are anticipated. 

\subsection{Undrained loading-unloading-reloading tests}

In addition to the calibration tests described above, undrained loading-unloading-reloading triaxial tests at zero cell pressure were carried out at axial strain rates of 160\,\%/hr with step relaxations, and at 16.0\,\%/hr, 1.6\,\%/hr, 4.81\,\%/hr and 0.16\,\%/hr for model validation. The test results will be presented together with model validation in Section\,\ref{sec:validation}.

\section{Parameter Fitting}
\label{sec:fits}

The hyperviscoplastic model proposed for undrained peat behaviour has 15 material parameters in total, viz. 8 hyperelastic material parameters ($C_1$, $D_2$ for the elastic response; $C_\mrm{1p}$, $D_\mrm{2p}$ for the elasto-plastic response; $C_\mrm{1v}^i$, $D_\mrm{2v}^i$ where $i = 1,\,2$ for the visco-elastic response); 4 parameters ($\alpha$, $\alpha_\mrm{p}$, $\alpha_\mrm{v}^i$) controlling the non-linearity of the constitutive relationship; one plastic parameter $c_\mrm{p}$; and 2 viscosity parameters $\eta_\mrm{v}^i$. For the undrained test conditions, the assumptions of incompressible and immiscible constituents of peat leads to a constant third principal invariant of the Right Cauchy Green tensor $I_3 \approx 1.0$. In the numerical implementation of the constitutive model, material deformation is the solution of the global Newton-Raphson iterations. The strain-controlled load and cell pressure of 0\,kPa represent the boundary conditions in the experimental triaxial tests performed. Near incompressibility is achieved numerically by setting the bulk modulus-like parameters $D_2$ to a very high value, thus penalizing volumetric deformation. The $D_2$ parameters were thus not subjected to fitting.

\subsection{Experimental Data}

To fit the numerical simulation to the experimental data, both need to be expressed in the same stress and strain spaces. For the experimental data, the stress measure was Cauchy stress $\mbf{\sigma}$, i.e. force per current area, and the recorded strain was engineering strain $\varepsilon_\mrm{eng} = \Delta h/h_0$; whereas the numerical data are expressed in terms of second Piola-Kirchhoff stress $\mbf{S}$ and Green-Lagrange strain $\mbf{E}$, respectively. The experimentally measured Cauchy stress in the specimen's axial direction equals the differential stress ($q = \sigma_{11} - \sigma_{33}$), i.e. $\sigma_{11} = q$, given a cell pressure of $\sigma_{33} = 0\,\mrm{kPa}$. The deformation gradient for the undrained triaxial compression can be obtained from  the vertical compression $\lambda = 1 + \varepsilon_\mrm{eng}$ as

\begin{equation} \label{eq:DFundrained}
 \mbf{F} = \left( \begin{array}{ccc}
 \lambda & 0 & 0\\
 0 &\frac{1}{\sqrt{\lambda}} & 0\\
 0 & 0 & \frac{1}{\sqrt{\lambda}}
\end{array}  \right)
\end{equation}

\noindent The Cauchy stress and the second Piola-Kirchhoff stress are related via 

\begin{equation}
\mbf{S} = J \mbf{F}^{-1} \mbf{\sigma} \mbf{F}^\mrm{-T} \label{eq:sigmaS}
\end{equation}

\noindent where $J$ is the determinant of the deformation gradient tensor, with a value of unity for the undrained test condition, cf. Eq.~\eqref{eq:DFundrained}. Following this post-processing of the experimental data, model parameters were fitted by comparing the numerical simulation results with the experimental data. 

\subsection{Equilibrium State Test}

Material parameters for the rate-independent part of the model were first fitted to the equilibrium test data. Thus the hyperviscoelastic components of the proposed model were eliminated by setting the $i^\mrm{th}$ viscoelastic parameters to vanishing values. To facilitate the fitting, the material parameter sensitivities of the constitutive relationships were studied. 


The experimental equilibrium test was simulated by prescribing boundary conditions as 0\,kPa cell pressure and an axial strain rate of 0.16\,\%/hour compressing the test specimen to an axial engineering strain of 20\,\%, followed by unloading at the same rate. When the stress of the axial direction reached zero during the unloading, the strain-controlled load case switched to stress-control maintaining the vertically stress-free state. Material parameter sensitivity was tested with the hyperelastoplastic parameters listed in Table \ref{tab:paraEP} and the results are plotted against the corresponding experimental results in Fig.~\ref{fig:para_ep}.

\begin{table*}[th!]
\centering
\caption{Parameter sensitivity test of the hyperelastoplastic model.}
\label{tab:paraEP}
\small
\begin{tabular}{c|cccccc}
\hline
 & $C_1$ [kPa] & $D_2$, $D_\mrm{2p}$ [kPa] & $\alpha$ & $C_\mrm{1p}$ [kPa] & $\alpha_\mrm{p}$ & $c_\mrm{p}$ [kPa\uexp{-1}]\\
\hline
Set $C_1$ & 9.0$\pm$5.0 & 500.0 & 0.0 & 50.0 & 0.0 & 0.1\\
Set $D_2 / D_\mrm{2p}$ & 9.0 & 1000.0$\pm$ 500.0 & 0.0 & 50.0 & 0.0 & 0.1\\
Set $\alpha$ & 9.0 & 500.0 & 0.5$\pm$0.5 & 50.0 & 0.0 & 0.1\\
Set $C_\mrm{1p}$ & 9.0 & 500.0 & 0.0 & 50.0$\pm$20.0 & 0.0 & 0.1\\
Set $\alpha_\mrm{p}$ & 9.0 & 500.0 & 0.0 & 50.0 & 100.0$\pm$100.0 & 0.1\\
Set $c_\mrm{p}$ & 9.0 & 500.0 & 0.0 & 50.0 & 0.0 & 0.1$\pm$0.02\\
\hline
\end{tabular}
\end{table*}


 %

\newcommand\Image[3][]{
 \centering 
 \tabular[b]{@{}c@{}}
 \includegraphics[#1]{#2}\\
 #3
 \endtabular}

{\Image[scale = 0.365]{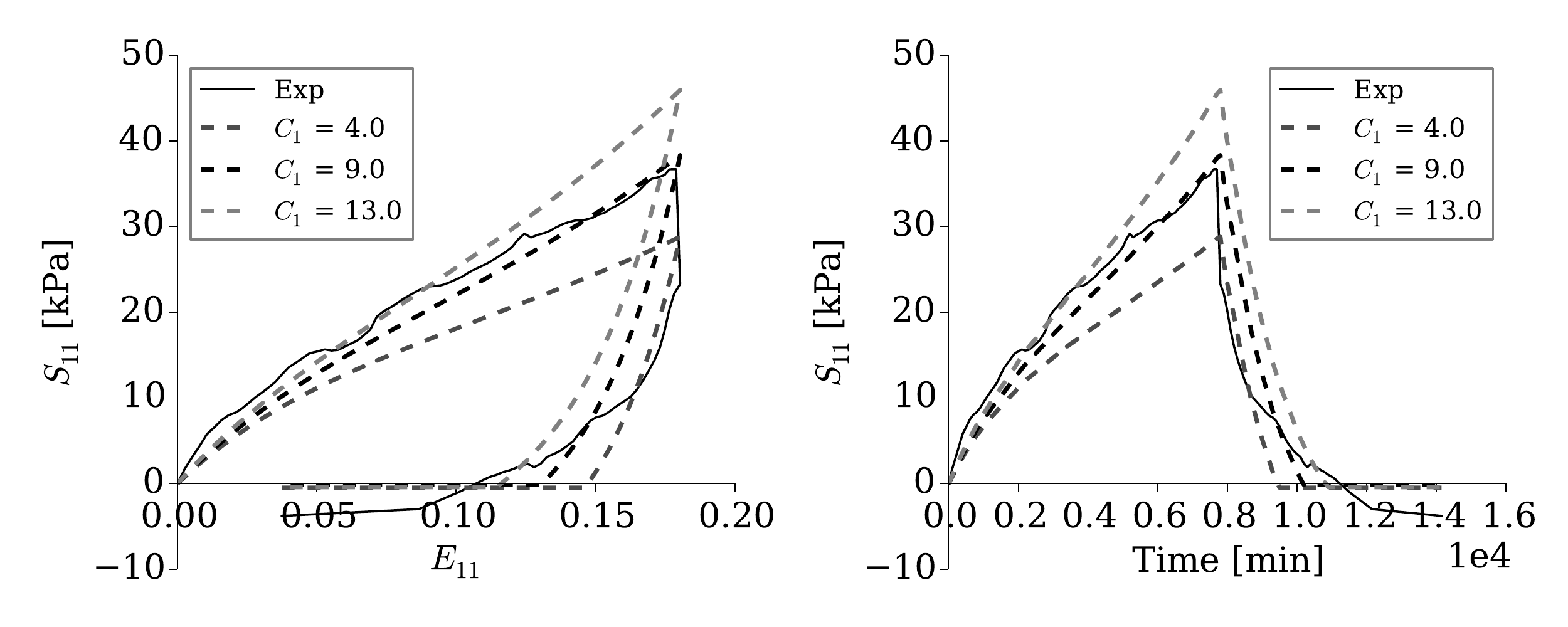}{(a) Set $C_\mrm{1}$.}\,
\Image[scale = 0.365]{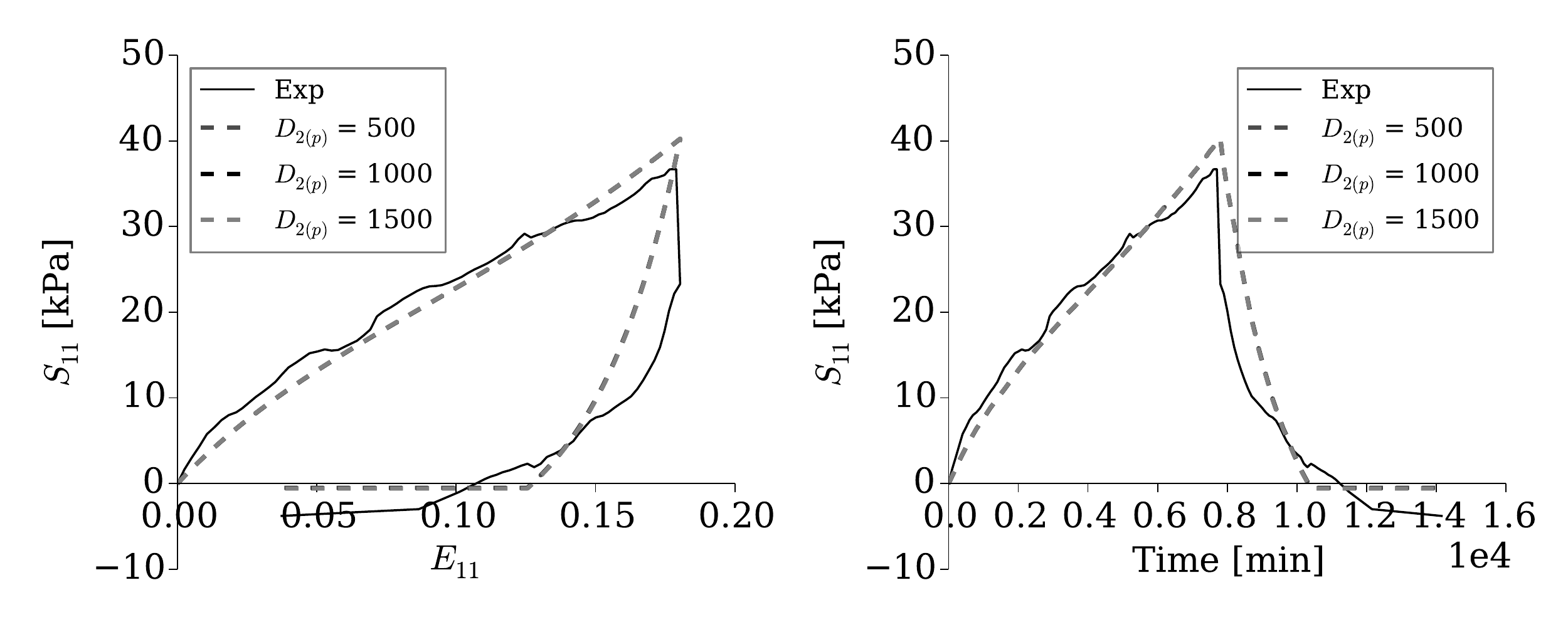}{(b) Set $D_2$ and $D_\mrm{2p}$.}\,
\Image[scale = 0.365]{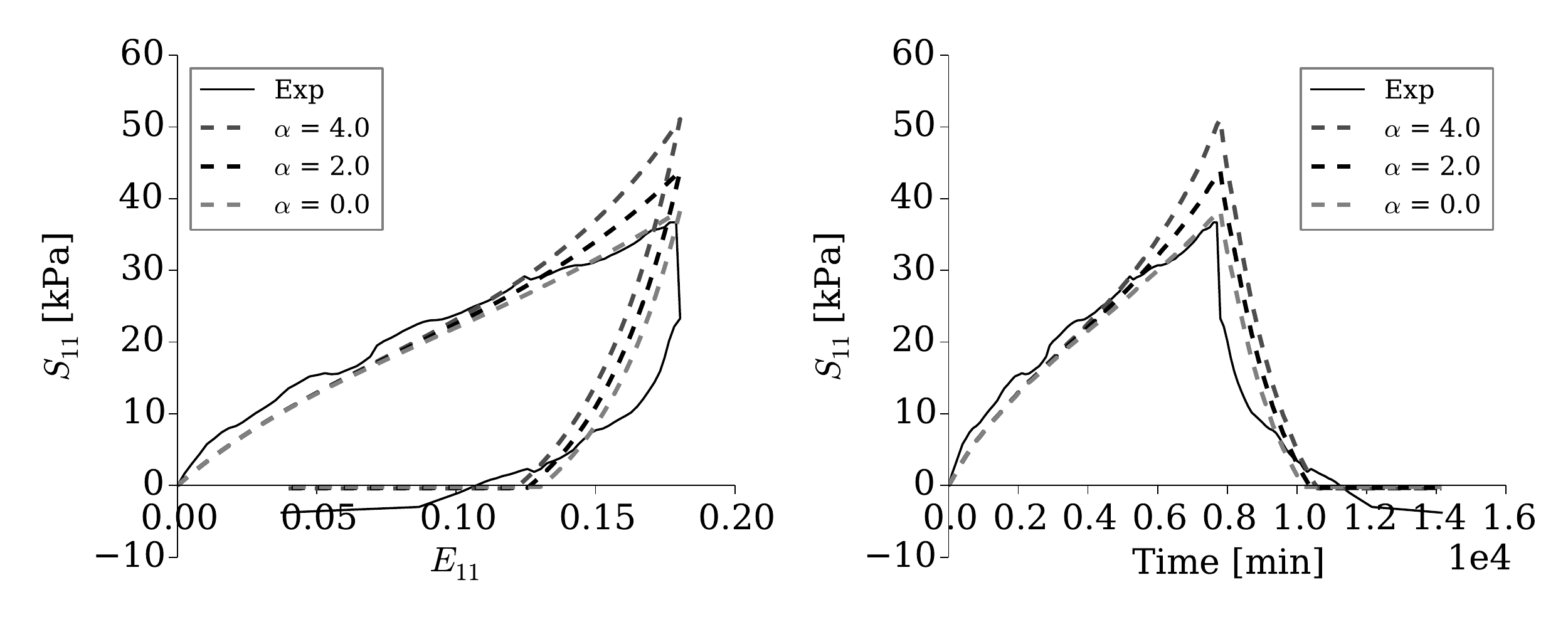}{(c) Set $\alpha$.}\,
\Image[scale = 0.365]{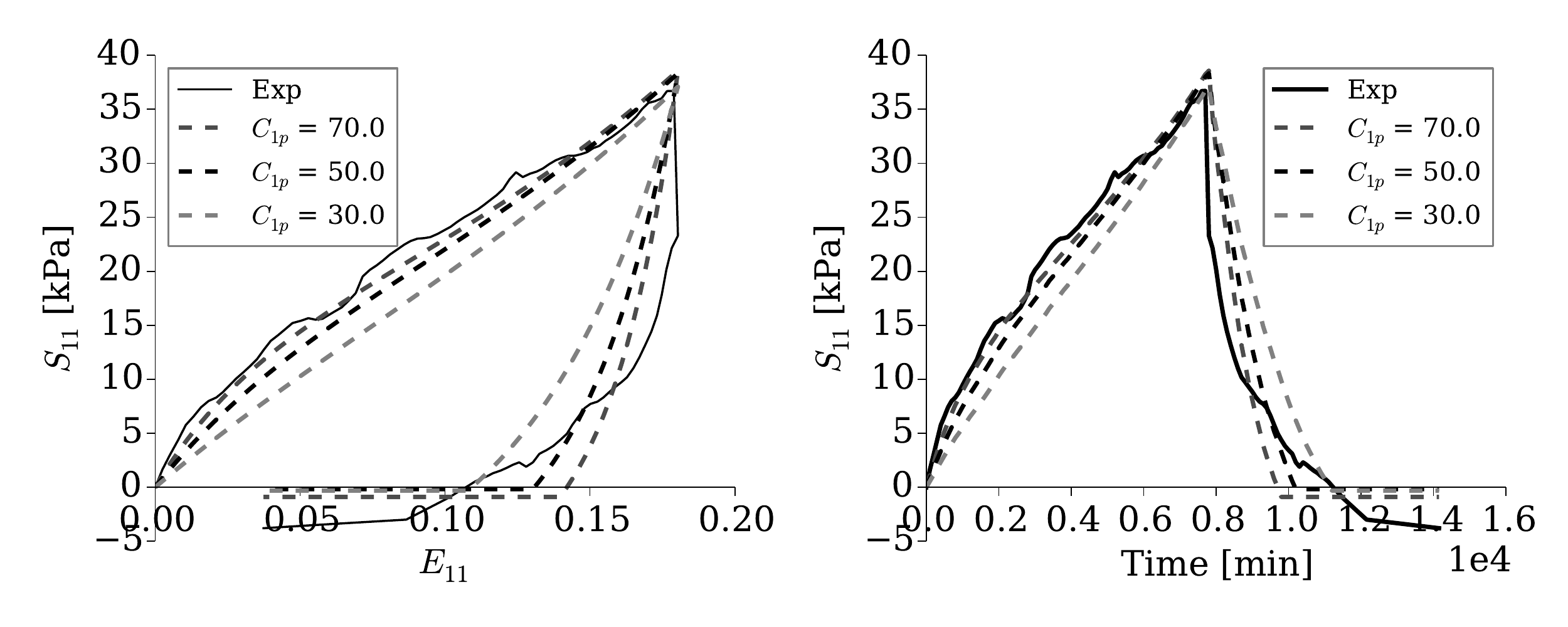}{(d) Set $C_\mrm{1p}$.}\,
\Image[scale = 0.365]{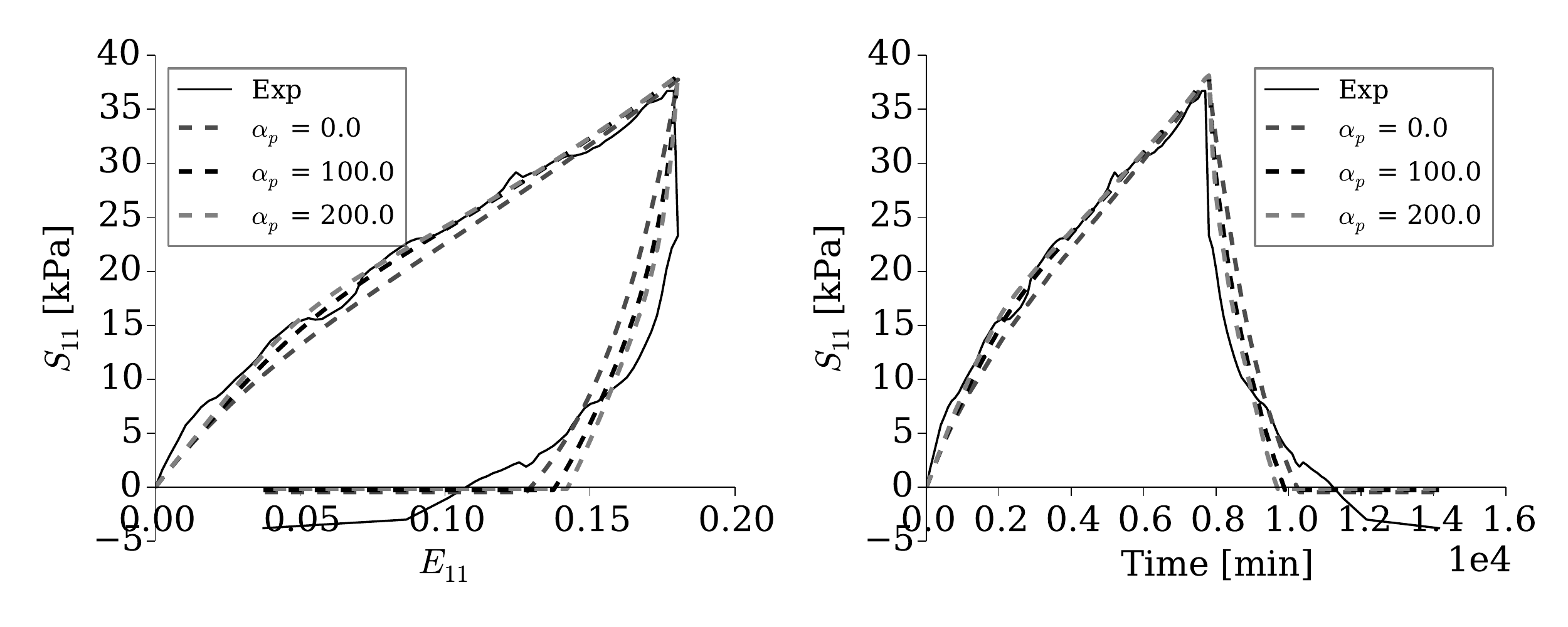}{(e) Set $\alpha_\mrm{p}$.}\,
\Image[scale = 0.365]{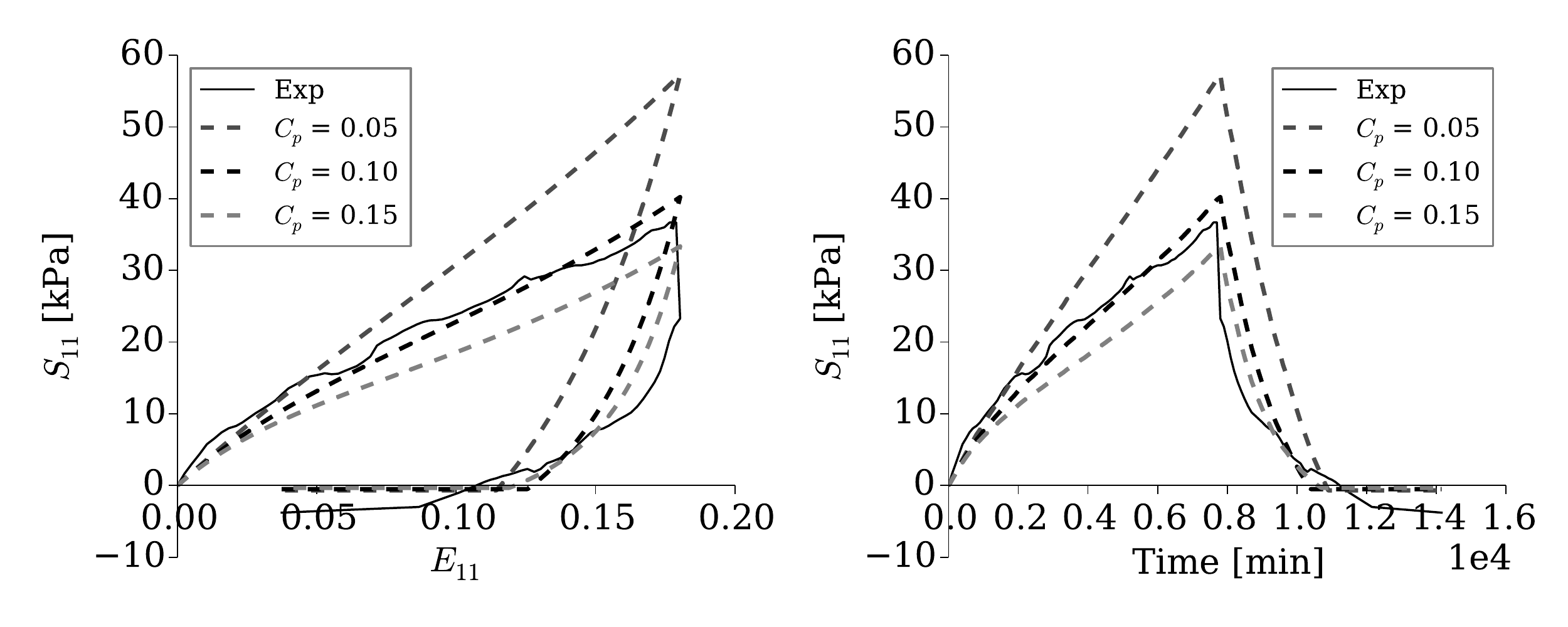}{(f) Set $c_\mrm{p}$.}
\captionof{figure}{Parameter sensitivity tests of the hyperelastoplastic model.} \label{fig:para_ep}}

Assigned with a relatively large value (500\,kPa), parameters $D_2$ and $D_\mrm{2p}$ had no influence on the constitutive relationship and successfully maintained the isochoric nature of the tests. Accordingly, the third invariant of the right Cauchy-Green deformation tensor in the simulation remained at 1.0$\pm$0.005. The loading curve was a hybrid effect of the parameters $C_1$, $C_\mrm{1p}$, $c_\mrm{p}$, $\alpha$ and $\alpha_\mrm{p}$. A higher $C_1$ indicated a larger loading stiffness and a smaller irrecoverable plastic strain. $C_\mrm{1p}$ determined the elastoplastic stress, thus influencing the plastic strain via the plastic flow rule. A larger $c_\mrm{p}$ transformed a greater amount of the elastoplastic stress into plastic strain. The shape parameters $\alpha$ and $\alpha_\mrm{p}$ had the same effect on the elastic spring and the elastoplastic spring elements, where $\alpha$ had a larger impact on the nonlinearity of the constitutive relationship with the selected material parameters.

\begin{figure}[th!]
\begin{center}
\includegraphics[scale=0.365]{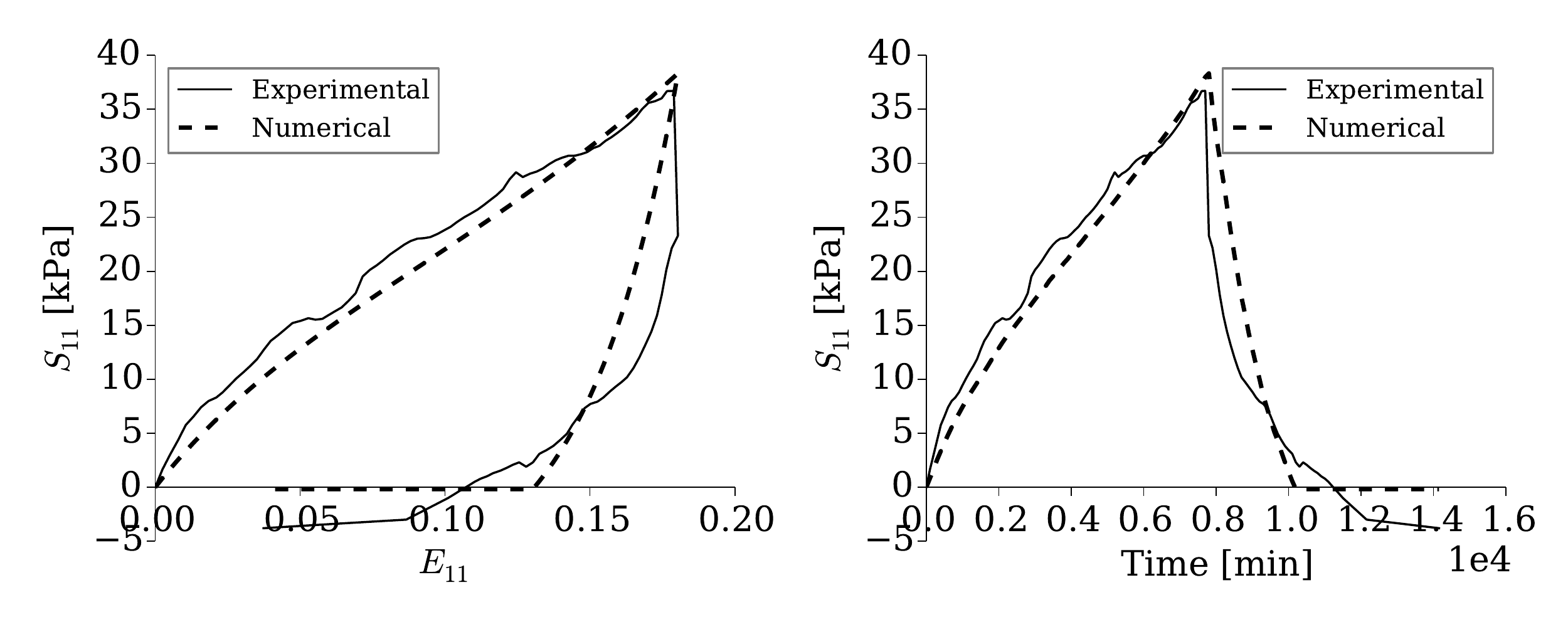}
\caption{The result of the hyperelastoplastic model fit to the equilibrium test of undrained peat.} \label{fig:ch6_para_ep_bstfit}
\end{center}
\end{figure}

The objective of the material parameter fitting was to find the proper combination of these parameters to balance between the loading curve and the irrecoverable plastic strain during unloading. The curve fitting process was carried out manually. Based on the hyperelastoplastic model parameter sensitivity tests, a set of the elastoplastic parameters was chosen that satisfactorily fulfilled this requirement. This set is presented in Table\,\ref{tab:para_ep} with the model fitting results shown in Fig.~\ref{fig:ch6_para_ep_bstfit}.

\begin{table}[th!]
\centering
\caption{Material parameters of the hyperelastoplastic model for the tested peat in undrained condition.}
\label{tab:para_ep}
\small
\begin{tabular}{ccccccc}
\hline
$C_1$ [kPa] & $D_2$ [kPa] & $\alpha$ & $C_\mrm{1p}$ [kPa]& $D_\mrm{2p}$ [kPa] & $\alpha_\mrm{p}$ & $c_\mrm{p}$ [kPa\uexp{-1}]\\
\hline
9.0 & 500.0 & 0.0 & 50.0 & 500.0 & 0.0 & 0.1\\
\hline
\end{tabular}
\end{table}

\subsection{Rate-dependent Tests}

\begin{figure}[th!]
\begin{center}
\includegraphics[scale=0.365]{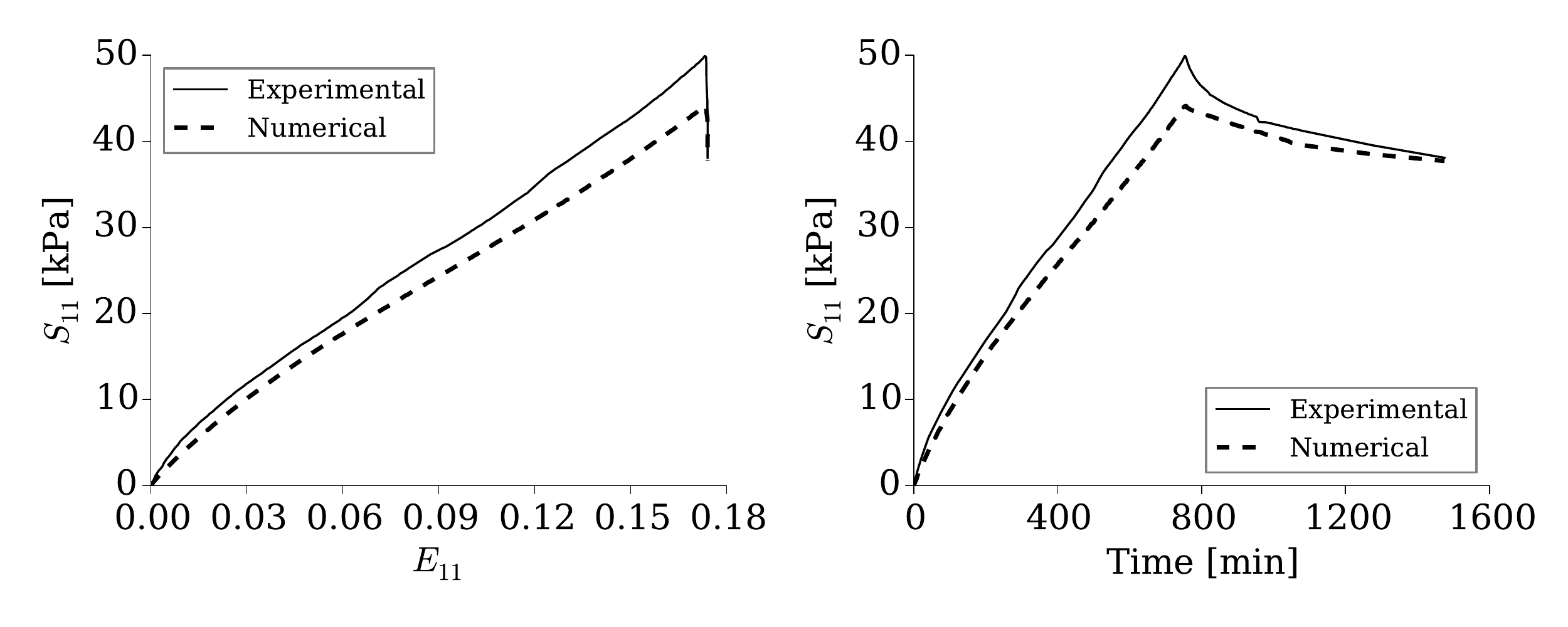}
\caption{Result of the full hyperviscoplastic model fit to an undrained triaxial compression-relaxation test at 1.6\,\%/hour compression rate.} \label{fig:ch6_para_vep_fitting1}
\end{center}
\end{figure}

\begin{figure}[th!]
\begin{center}
\includegraphics[scale=0.365]{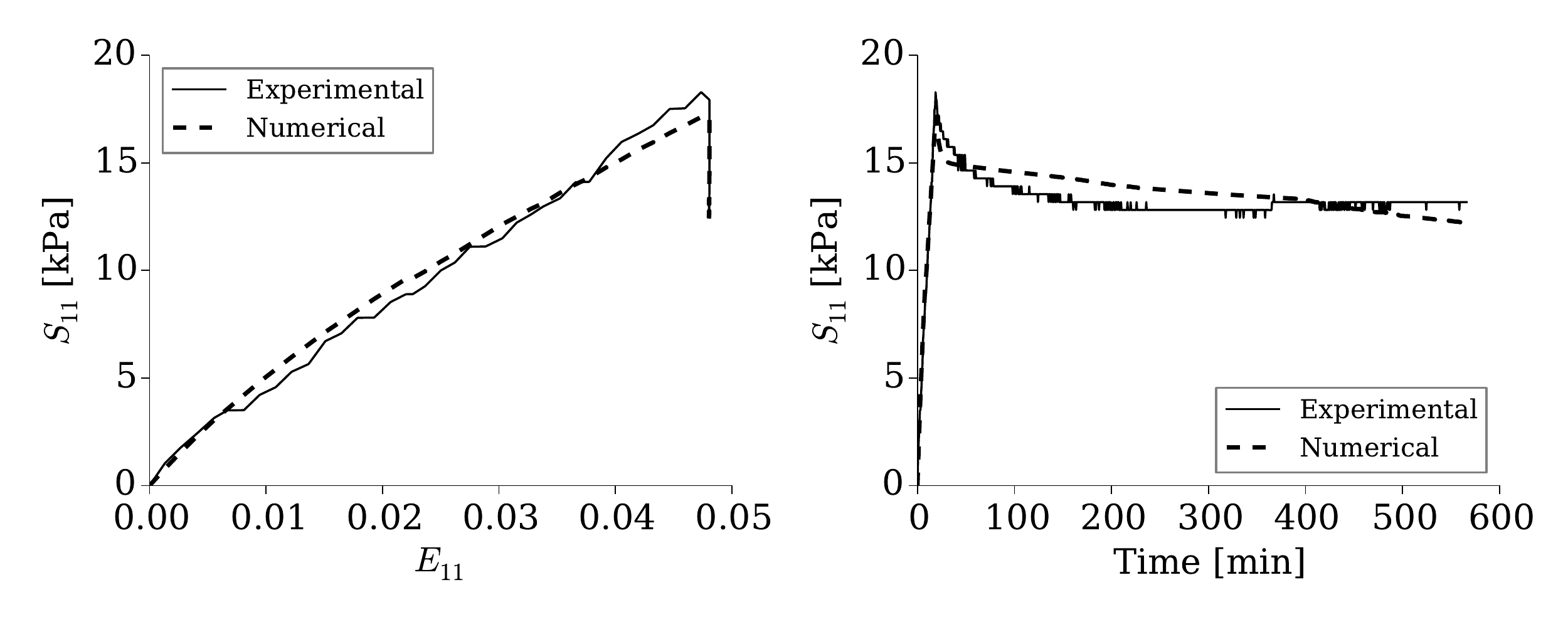}
\caption{Result of the full hyperviscoplastic model fit to an undrained triaxial compression-relaxation test at 16.0\,\%/hour compression rate.} \label{fig:ch6_para_vep_fitting2}
\end{center}
\end{figure}

Freezing the fitted material parameters for the hyperelastoplastic model, tests at higher strain rates are needed for fitting the rate-dependent parts of the model. Undrained triaxial compression tests at two strain rates with relaxation were used for the parameter fitting of the two rate-dependent hyperviscoelastic model components. As outlined earlier, the parameters $D_\mrm{2v}^1$ and $D_\mrm{2v}^2$ have no influence on the simulated constitutive relationships under the present considerations, and were thus assigned with same values as $D_2$ and $D_\mrm{2p}$ in the hyperelastoplastic component. Analogous to the hyperelastoplastic model, $\alpha_\mrm{v}^1$ and $\alpha_\mrm{v}^2$ are assigned to be zero, simplifying the Helmholtz free energy function \eqref{eq:psiv} to a Neo-Hookean model. The model parameters $C_\mrm{1v}$ and $\eta_\mrm{v}$ of each viscoelastic component obtained by fitting the undrained triaxial compression tests with relaxation at the compression strain rates of 1.6\,\%/hour and 16.0\,\%/hour are given in Table\,\ref{tab:para_ev}. The results are presented in Figs.~\ref{fig:ch6_para_vep_fitting1} and \ref{fig:ch6_para_vep_fitting2}, respectively. Experimental and numerical results show some deviations but agree at least qualitatively to a satisfactory level considering the inter-specimen variations for peat materials and the temperature effect experienced during the experimental tests.

\begin{table}[th!]
\centering
\caption{Material parameters of the hyperviscoelastic model for the undrained peat.}
\label{tab:para_ev}
\small
\begin{tabular}{c|cccc}
\hline
Parameter & $C_\mrm{1v}^i$ [kPa] & $D_\mrm{2v}^i$ [kPa] & $\alpha_\mrm{v}^i$ &  $\eta_\mrm{v}^i$ [kPa\,d]\\
\hline
$i=1$ & 8.0 & 500.0 & 0.0 & 9.0\\
$i=2$ & 40.0 & 500.0 & 0.0 & 0.35\\
\hline
\end{tabular}
\end{table}

\section{Model Validation} \label{sec:validation}

The hyperviscoplastic model with material parameters from Tables \ref{tab:para_ep} and \ref{tab:para_ev} was compared against undrained triaxial tests performed on five undisturbed vertical peat specimens at axial engineering strain rates of 160.0, 16.0, 4.81, 1.60 and 0.16\,\%/hour, which had not been used previously for fitting the model parameters. The strain-controlled loads from the experiments were directly applied to the numerical simulations. To compare the proposed model performance with the experimental results in both compression and unloading stages, the load cases for the following validations were in an utterly strain-controlled manner, thus tensile stresses were created during unloading as no platen lift-off was allowed. 



{\Image[scale = 0.365]{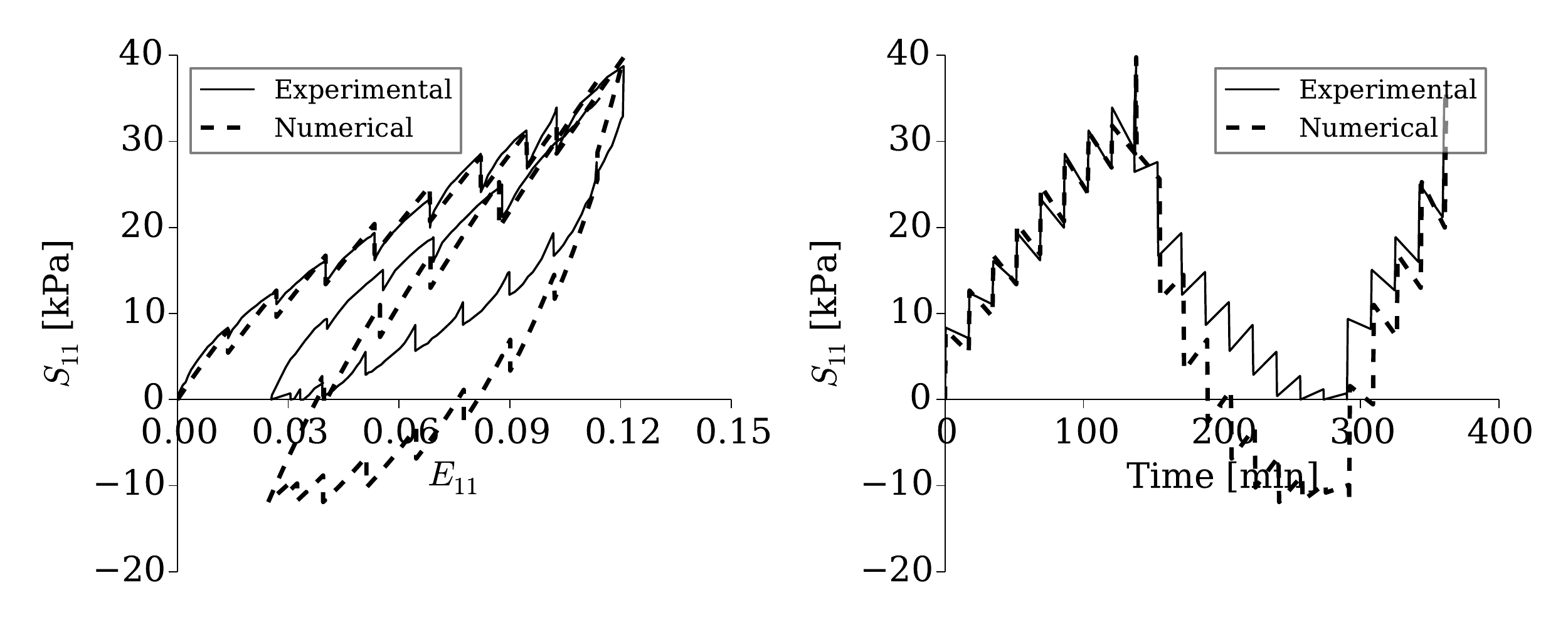}{(a) 160.0\,\%/hour.}\,
\Image[scale = 0.365]{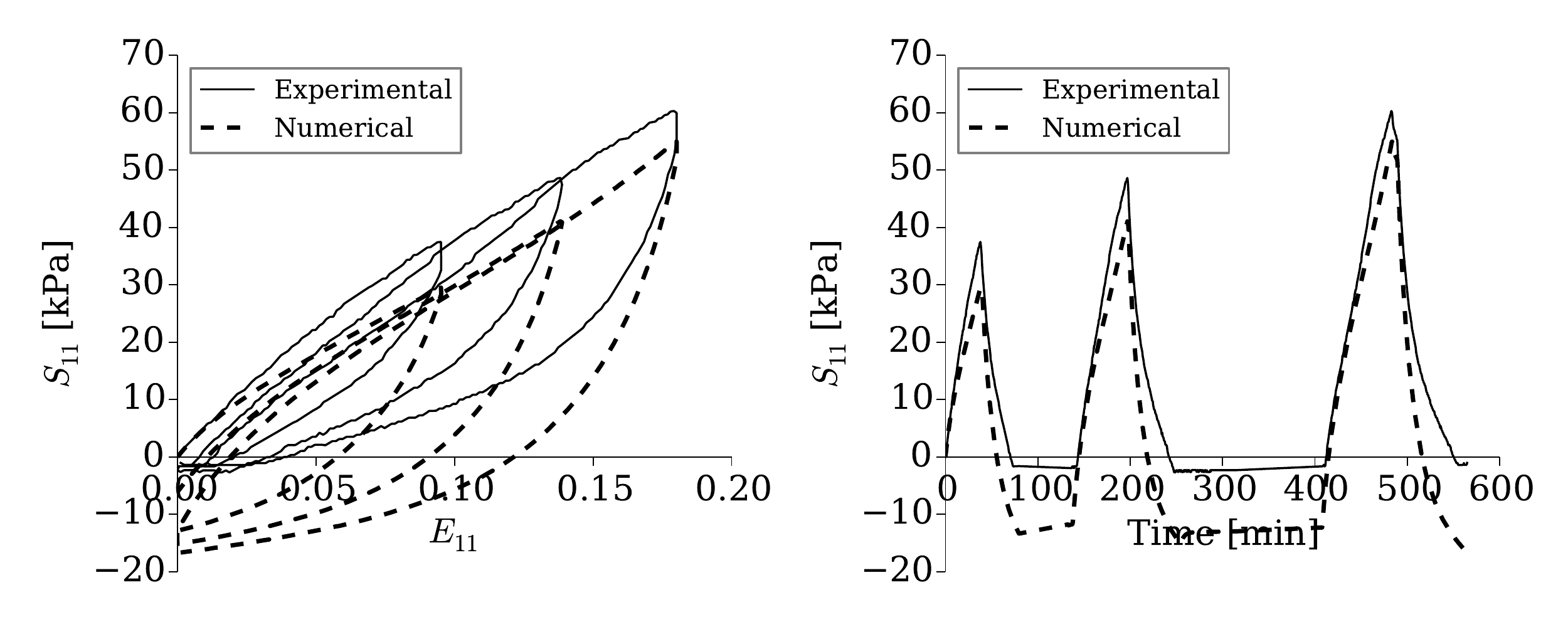}{(b) 16.0\,\%/hour.}\,
\Image[scale = 0.365]{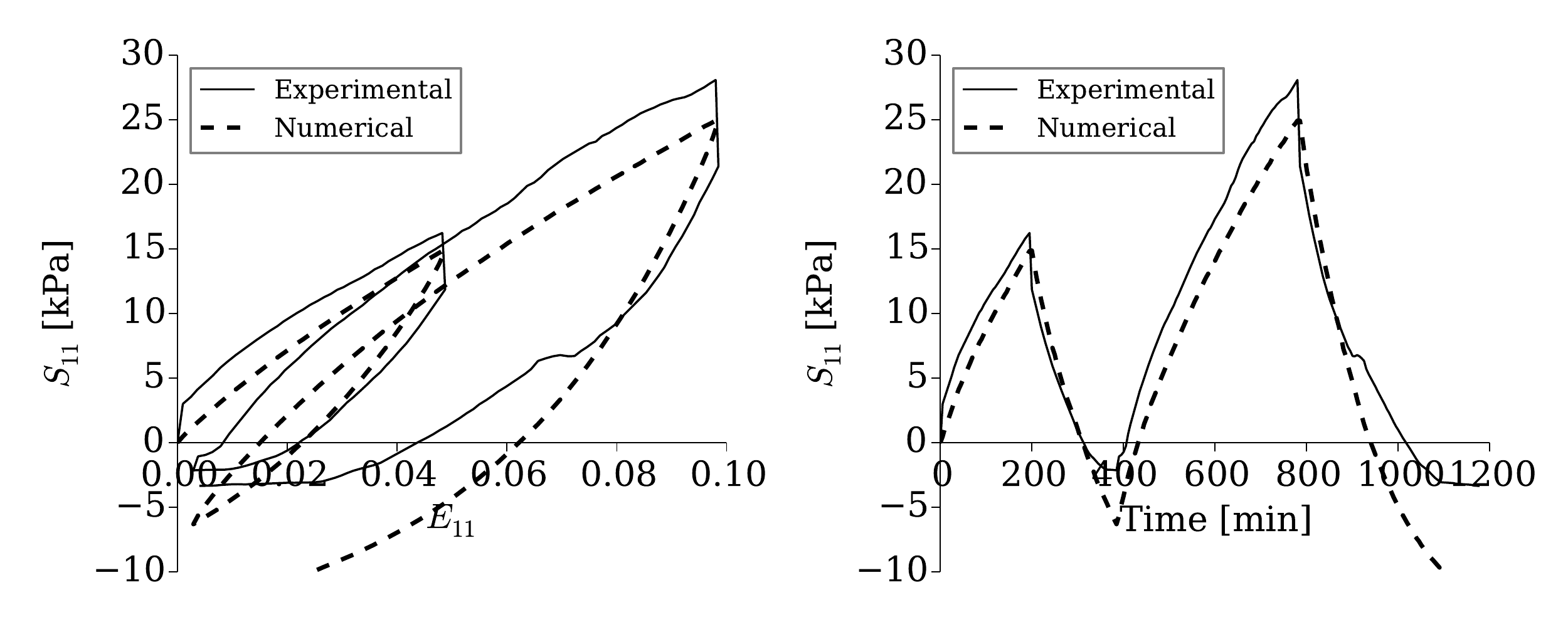}{(c) 1.6\,\%/hour. }\,
\Image[scale = 0.365]{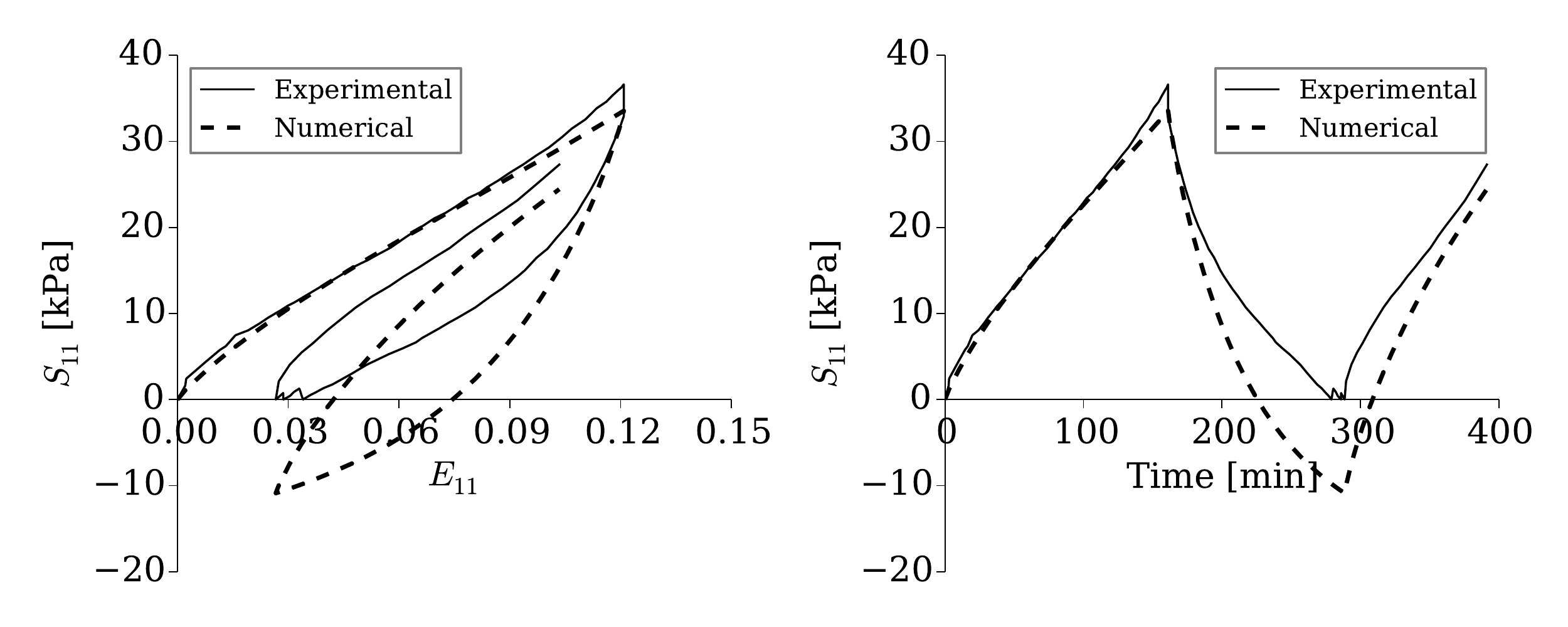}{(d) 4.81\,\%/hour.}\,
\Image[scale = 0.365]{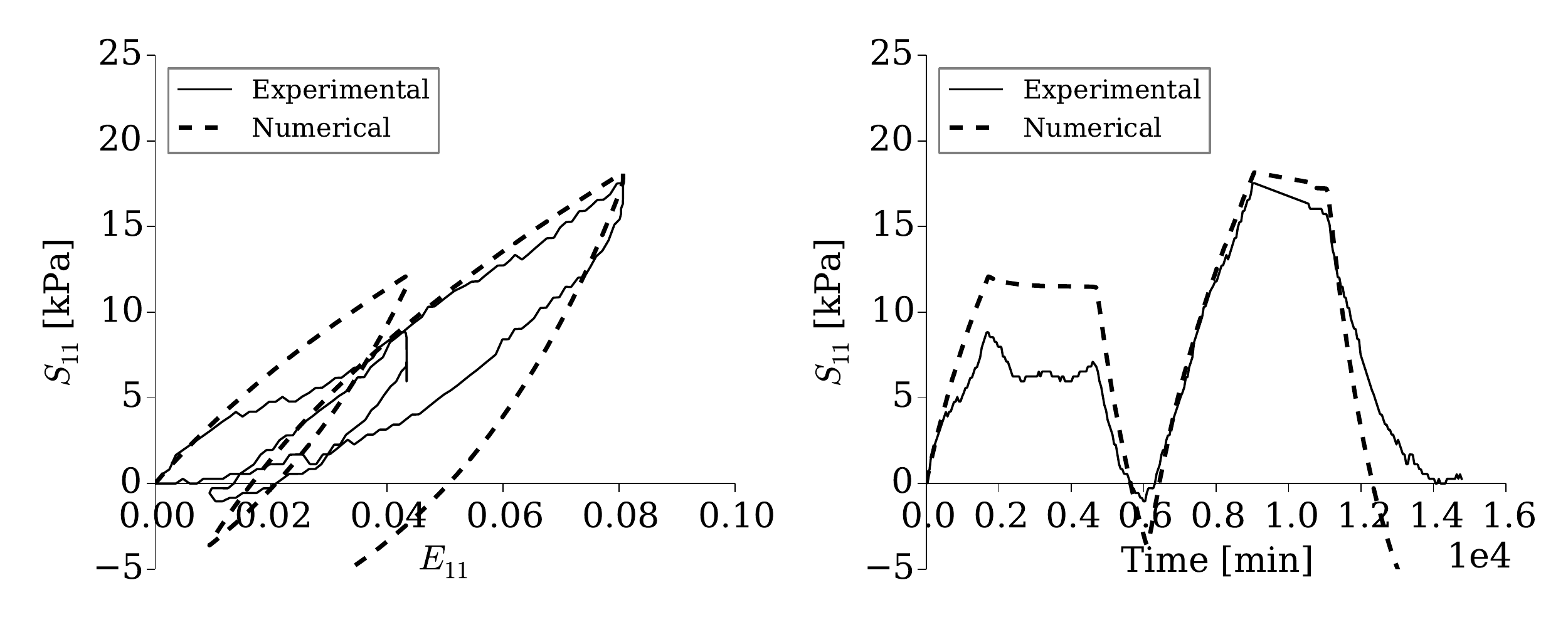}{(e) 0.16\,\%/hour.}
\captionof{figure}{Hyperviscoplastic model validations.} \label{fig:evp_validation}}

The experimental data presented along with the model validation demonstrated the challenges in testing peat, viz. the intrinsic specimen variability, the manifest sensitivity on load measurements and temperature fluctuations due to the relatively low stress level. The equilibrium irrecoverable strains in the loading-unloading cycles to an axial strain of 20\,\% range from 2.8\,\% to 8.5\,\%. The hyperelastoplastic model parameters were obtained by fitting to the equilibrium loading-unloading test with a relatively large irrecoverable strain (8.5\,\%), thus resulting in a significant discrepancy between the simulation and the experimental results in unloading for the other tests which showed a quantitatively quite different hysteresis behaviour. Taking into account this large inter-specimen variability and the temperature effect, the simulation results in Fig.~\ref{fig:evp_validation} show that the proposed hyperviscoplastic model was able to at least qualitatively predict peat behaviour in undrained triaxial compression tests for five different strain rates with step relaxations to which it was not fitted. This was true even for strain-rate ranges outside of the fitting domain. To test the unloading performance of the proposed model, the model was allowed to generate tensile stress at the end of unloading as the strain-controlled load was continuously applied albeit the stress reached negative values, whereas in the laboratory experiments, the load cell detached from the test specimens. Some initial remediation of this effect will be indicated Section~\ref{sec:disc}.

\section{Discussion}
\label{sec:disc}

The validation of the hyperviscoplastic model demonstrated the model's principal capability to predict other tests in a semi-quantitative fashion, with a better performance during the loading stage of the undrained triaxial tests. The relatively larger discrepancy during unloading between the numerical simulation and the experimental results could be improved by correcting the load case and modifying the plastic flow rule. 

\subsection{Load case modification}

\begin{figure}[th!]
\begin{center}
\includegraphics[scale=0.365]{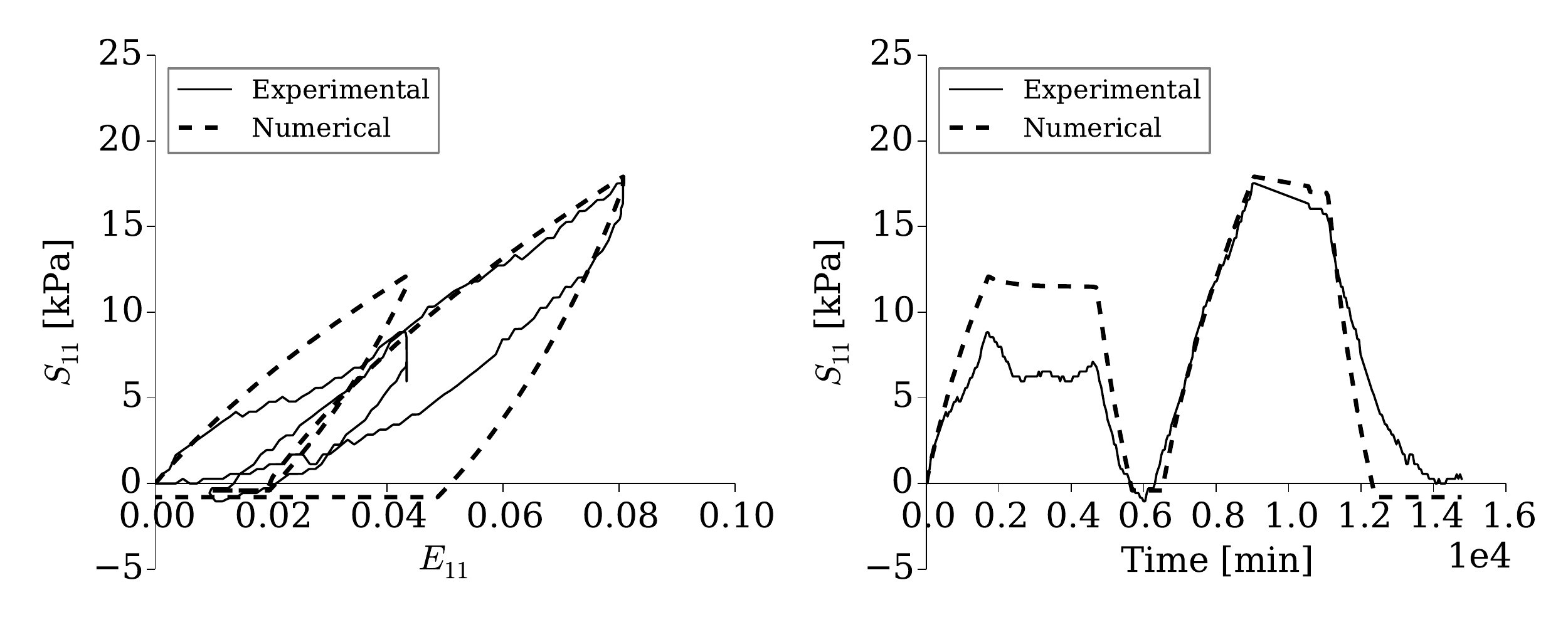}
\caption{Hyperviscoplastic model behaviour with modified load case compared to the equlibrium test of undrained peat.} \label{fig:ch6_para_vep_1V6_lc2}
\end{center}
\end{figure}

The load case modification was realized by allowing a detachment of the loading platen from the material. During the separation period between the loading platen and the material, the material was allowed to creep under a vanishing applied stress. When the loading platen reached the position of the creeping material during reloading, the strain-controlled load was reactivated. The hyperviscoplastic model with the modified load case is re-validated against the equilibrium test in Fig.~\ref{fig:evp_validation}\,(e). Fig.~\ref{fig:ch6_para_vep_1V6_lc2} presents the modified load case which better reflected the test loading conditions.

\begin{figure}[th!]
\begin{center}
\includegraphics[scale=0.365]{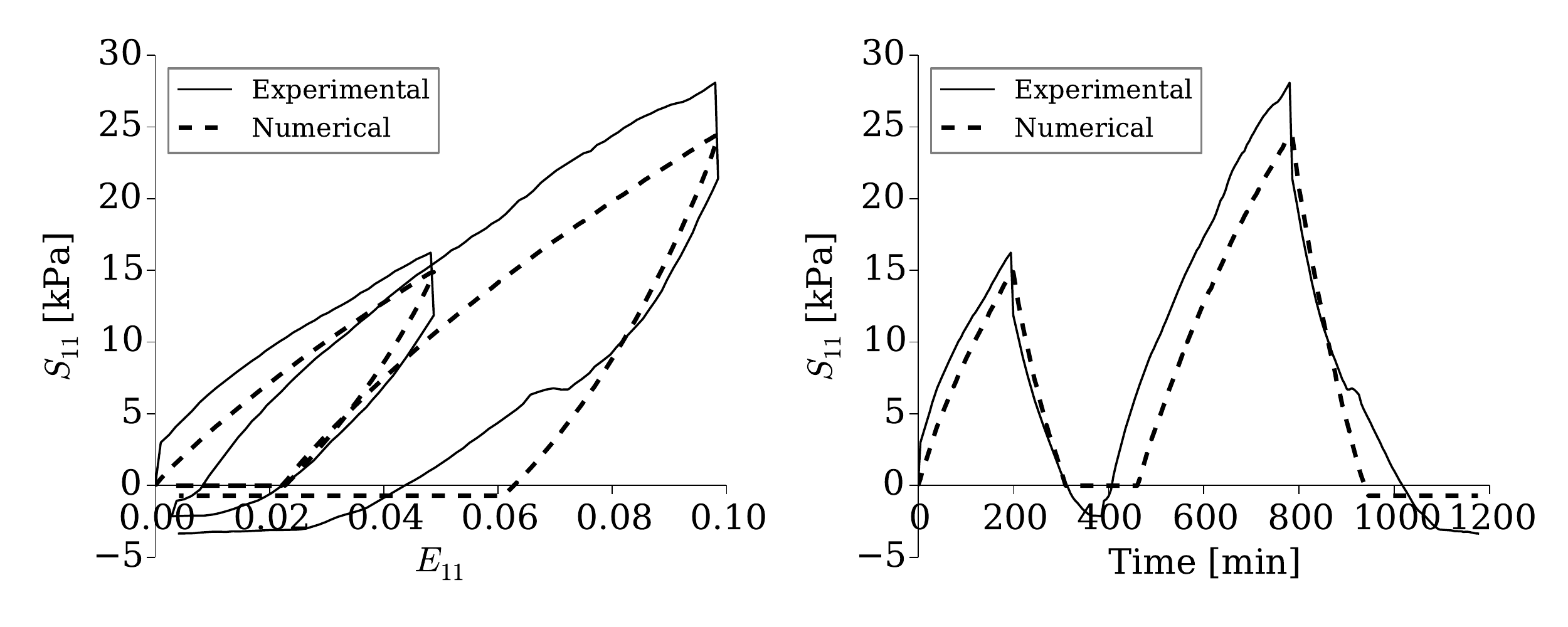}
\caption{Hyperviscoplastic model behaviour with modified load case compared to the undrained triaxial test at 1.6\,\%/hour.} \label{fig:ch6_para_vep_2V8lv2}
\end{center}
\end{figure}

As the proposed model was capable of simulating tests carried out at different strain rates, the triaxial loading-unloading-reloading test shown in Fig.~\ref{fig:evp_validation}\,(c) was used as a rate-dependent representative test with strain-cycles for further investigation of the behaviour under the modified load case. The simulation result is presented in Fig.~\ref{fig:ch6_para_vep_2V8lv2} and the unloading tensile stress was banished by the modified load case. The reloading curve was shifted to a larger irrecoverable plastic strain from the former unloading.

\subsection{Plastic flow rule modification}

The plastic flow rule was modified by eliminating the plastic strain evolution during passive unloading. When the applied load switched from loading to unloading, the material was still within the compressive domain. The original plastic flow rule in Eq.~\eqref{eq:frp} determines the plastic strain increment direction as co-aligned with the current stress in the plastic element. Therefore, a modification of the flow rule was performed by suppressing plastic flow during passive unloading. Plastic flow appeared again during active load reversal. The modification was realized by multiplying a Heaviside step function to the current plastic flow rule that activates flow only if strain evolution and the stress in the plastic element driving the flow are aligned in the sense of a norm. Specifically, the current plastic flow rule \eqref{eq:intfrp} was modified in the intermediate configuration to

\begin{equation} \label{eq:plasticflowrule_modify} 
 \overset{\Delta}{\mbf{\epsilon}}_\mrm{p} := c_\mrm{p} \rho_0 || \overset{\Delta}{\mbf{\epsilon}}||\left( \mbf{I} + 2 \mbf{\epsilon}_\mrm{ep} \right) \frac{\partial \bar{\psi}_\mrm{p}}{\partial \mbf{\epsilon}_\mrm{ep}} H\left[ \mrm{sign} \left( \rho_0 \frac{\partial \bar{\psi}_\mrm{p}}{\partial \mbf{\epsilon}_\mrm{ep}} \dcdot \overset{\Delta}{\mbf{\epsilon}} \right) \right]
\end{equation}

\noindent where $H(\bullet)$ is the Heaviside step function
and $\mrm{sign}(\bullet)$ is the signum function.
%

%

\begin{figure}[th!]
\begin{center}
\includegraphics[scale=0.365]{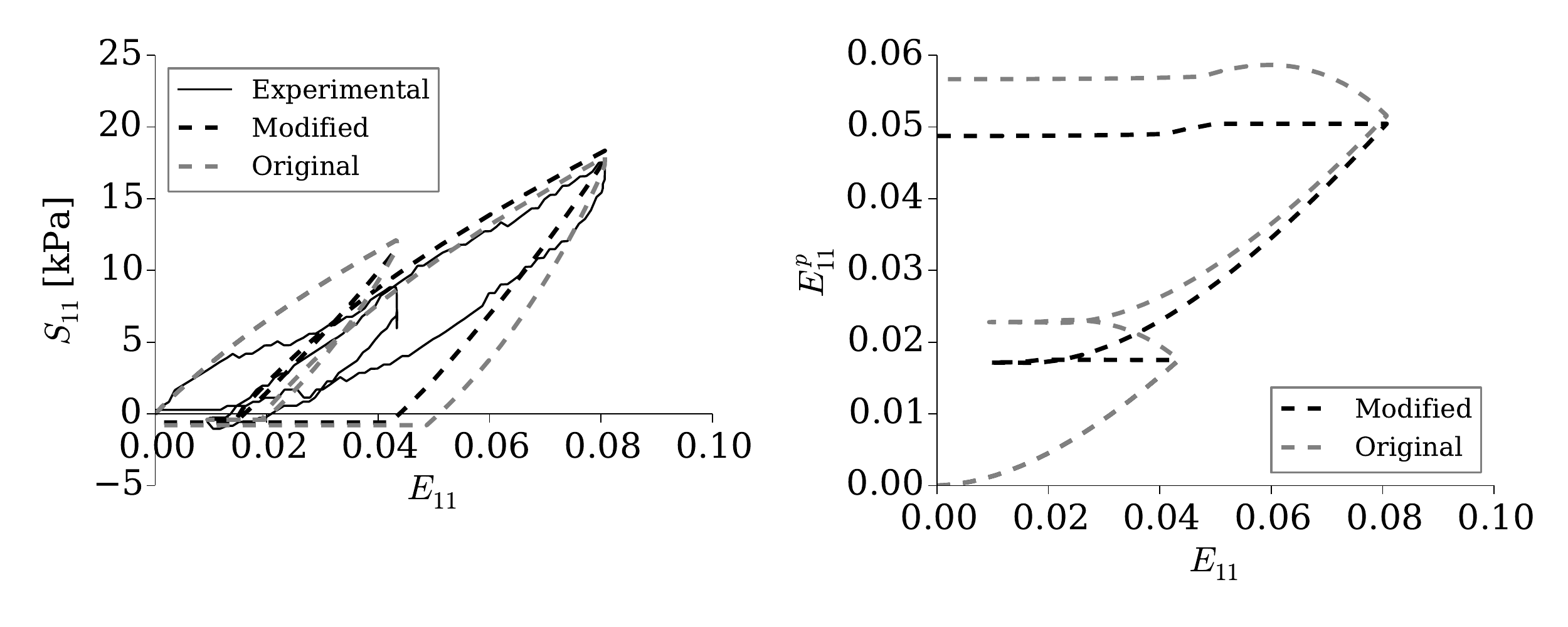}
\caption{Comparison of the original and modified plastic flow rule in representing the 0.16\,\%/hour test.} \label{fig:ch6_modifiedpf_equilibrium}
\end{center}
\end{figure}

\begin{figure}[th!]
\begin{center}
\includegraphics[scale=0.365]{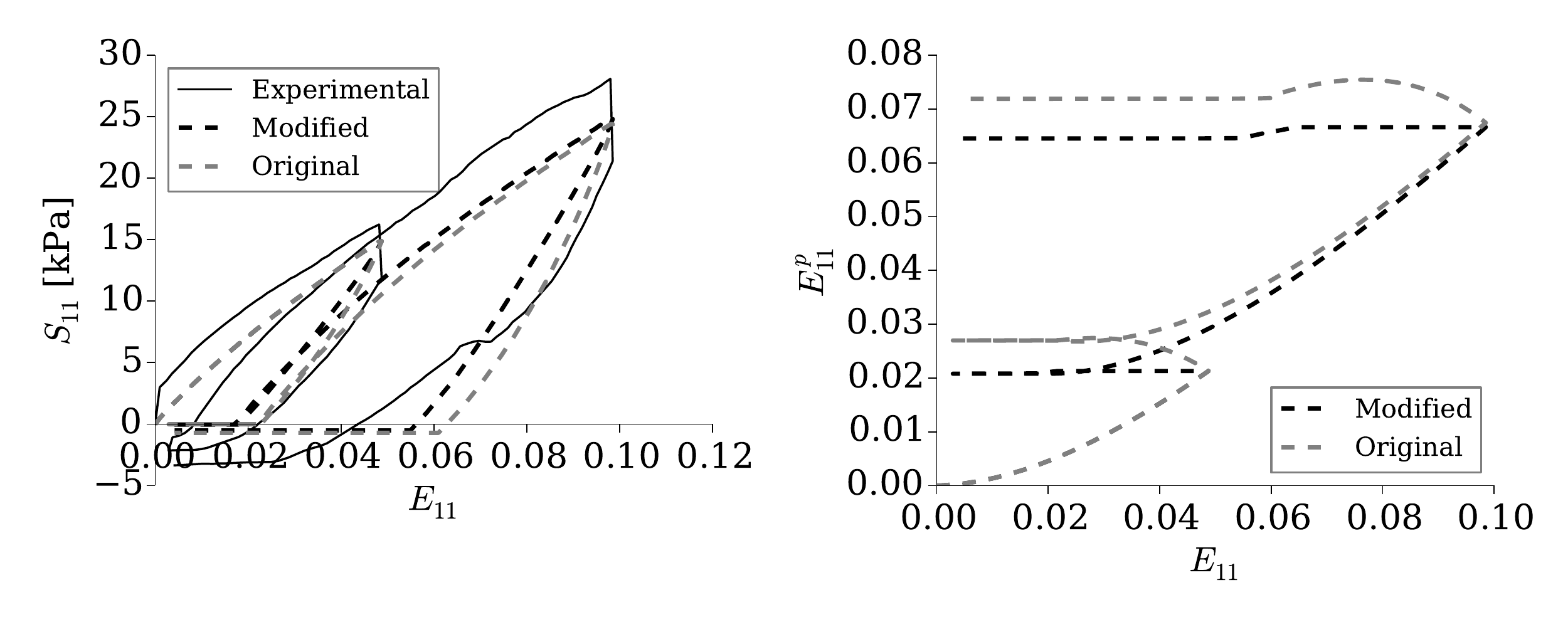}
\caption{Comparison of the original and modified plastic flow rule in representing the 1.6\,\%/hour test.} \label{fig:ch6_modifiedpf_2V8}
\end{center}
\end{figure}

The previously determined material parameters were used for the comparison of the hyperviscoplastic model with the modified plastic flow rule to the equilibrium test and the test at 1.6\,\%/hour. Fig.~\ref{fig:ch6_modifiedpf_equilibrium} and Fig.~\ref{fig:ch6_modifiedpf_2V8} present the simulation results of the tests carried out at 0.16\,\%/hour and 1.6\,\%/hour with the modified and the original flow rules. The partial plastic strains on the reference configuration, obtained by $\mbf{E}_\mrm{p} = \frac{1}{2}\left(  \mbf{C}_\mrm{p} - \mbf{I}\right)$, were compared. The plastic strain increase during the unloading compressive domain was eliminated and remained constant instead. The irrecoverable plastic strain of the constitutive relationship during unloading was lessened by the modified plastic flow rule bringing it closer to the experimentally observed values.

\section{Conclusions and Future Work}
\label{sec:concl}

Peat is a challenging natural material for both laboratory testing and numerical modelling. Among the outstanding features of this article in the context of the literature available on peat modelling, are the systematic derivation of a finite-strain constitutive model within a modern continuum mechanical thermodynamically consistent framework in combination with modelling-oriented laboratory experiments designed to extract information on flow-independent viscoelastic phenomena. A hyperviscoplastic model with its elastic, viscoelastic,and elastoplastic components connected in parallel was adopted for modelling peat. The model provided qualitatively good simulation results of the undrained compression of the tested peat and demonstrated its versatility in simulating different load cases (loading-unloading with step relaxations at different strain rates) as well as its potential for further developments. 

A certain advantage of a constitutive set-up like the presented one is its modular structure, with equilibrium and non-equilibrium parts modelled separately according to the experimental information available. Thus, certain "layers" of the model can simply be switched on or off depending on the modelling purpose and the availability of experimental data. The model can be improved in several ways, which are currently under study. 

Further experimental data with better control of laboratory conditions are required for peat, such as keeping temperature fluctuations within $\pm0.05$\,K. Additionally, a statistical investigation of the possible spread of sample-specific equilibrium states for this type of peat material should be proposed based on a relatively large quantity of tests. Specifically, the definition of the equilibrium hysteresis must be defined in a manner that is consistent with the hysteresis observed in higher-rate tests. The strain-energy functions and flow-rule definitions chosen here were selected due to their generic nature and their low number of parameters. This makes the model versatile but leaves room to devise energy functionals and flow rules tailored to the characteristics of peat. Such a step requires further experimental study but is expected to greatly improve the model's interpretive and predictive capabilities. Nevertheless, considering the generic nature of the expressions used here, the model already captured a wide spectrum of the experimental observations.

Furthermore, the model will be applied to peat under drained conditions. For that case, the various $D_2$-parameters governing the volumetric behaviour will have to be fitted in addition to quantifying flow-dependent effects by accounting for fluid-solid interaction. The model will, for that purpose, be included into a framework for hydromechanically coupled analyses, e.g. Nagel and Kelly \cite{Nagel2012} and G{\"o}rke et al. \cite{Gorke2010}, to capture transient effects linked to fluid pressurisation such as flow-dependent viscoelasticity. For that purpose, the devised model will constitute the constitutive law for the effective stresses $\mbf{S}_\mrm{S}^\mrm{E}$

\begin{equation}
	\mbf{S} = -p \mbf{C}^{-1} + \mbf{S}_\mrm{S}^\mrm{E}
\end{equation}

\noindent where the pore pressure $p$ is derived from a saturation-constrained mass balance equation of the porous medium extended by a flow-model, e.g. Darcy's law.

These extensions will include an implementation into a finite element framework to allow simulations with spatially heterogeneous solution and parameter fields. At that stage, the model will also be linked to advanced parameter identification algorithms to obtain material parameters by fitting to a series of tests simultaneously in an automated manner.

\section*{Acknowledgements}

The authors acknowledge Mr. Eoin Dunne and Dr. Kevin Ryan for their assistance in collecting the peat samples as well as setting up the laboratory testing. This research was funded by Postgraduate Embark Initiative (grant number RS/2011/271) from the Irish Research Council and a Postgraduate Studentship Award from Trinity College Dublin. The modelling work was performed during a research stay at the Helmholtz-Centre for Environmental Research - UFZ. We cordially thank Prof. Olaf Kolditz for hosting the first author during that period, and the School of Engineering, Trinity College Dublin, for providing a travel and subsistence allowance for the same.

\clearpage

\appendix

\section{Appendix: Local matrix-vector system}
\label{sec:apdx}


\noindent Residual vector of the numerical implementation for the hyperviscoplastic model:
		\begin{align}
& \begin{aligned}
 \mvec{r}_1 & = \widetilde{\mvec{S}} - \frac{2}{C_1} \left( \frac{\partial \psi_\mrm{0}}{\partial I_\mrm{1}^\mrm{eq0}} \mvec{I} + \frac{\partial \psi_\mrm{0}}{\partial I_\mrm{3}^\mrm{eq0}} I_\mrm{3}^\mrm{eq0} \mvec{C}^\mrm{-1} \right)- \frac{2}{C_1} \left( \frac{\partial \psi_\mrm{p}}{\partial I_\mrm{1}^\mrm{ep}} \mvec{C}_\mrm{p}^\mrm{-1} + \frac{\partial \psi_\mrm{p}}{\partial I_\mrm{3}^\mrm{ep}} I_\mrm{3}^\mrm{ep} \mvec{C}^\mrm{-1} \right)\\
 & -\sum_{i=1}^2 \frac{2}{C_1} \left( \frac{\partial \psi_\mrm{v}^i}{\partial I_\mrm{1}^{\mrm{ev}i}} \mvec{C}_{\mrm{v}i}^\mrm{-1} + \frac{\partial \psi_\mrm{v}^i}{\partial I_\mrm{3}^{\mrm{ev}i}} I_\mrm{3}^{\mrm{ev}i} \mvec{C}^\mrm{-1} \right)
 \end{aligned}\\
& \mvec{r}_2  = \frac{\mvec{C}_\mrm{v1} - \mvec{C}_\mrm{v1}^\mrm{t}}{\Delta t} - \frac{4}{\eta_\mrm{v1}} \left( \frac{\partial \psi_\mrm{v1}}{\partial I_1^\mrm{ev1}} \mvec{C} + I_3^\mrm{ev1} \frac{\partial \psi_\mrm{v1}}{\partial I_3^\mrm{ev1}} \mvec{C}_\mrm{v1} \right)\\
 & \mvec{r}_3  = \frac{\mvec{C}_\mrm{v2} - \mvec{C}_\mrm{v2}^\mrm{t}}{\Delta t} - \frac{4}{\eta_\mrm{v2}} \left( \frac{\partial \psi_\mrm{v2}}{\partial I_1^\mrm{ev2}} \mvec{C} + I_3^\mrm{ev2} \frac{\partial \psi_\mrm{v2}}{\partial I_3^\mrm{ev2}} \mvec{C}_\mrm{v2} \right)\\
 & \mvec{r}_4 = \frac{\mvec{C}_\mrm{p} - \mvec{C}_\mrm{p}^\mrm{t}}{\Delta t} - 2 c_p \frac{||\mvec{C} - \mvec{C}^\mrm{t}||}{\Delta t} \left( \frac{\partial \psi_\mrm{p}}{\partial I_1^\mrm{ep}} \mvec{C} + I_3^\mrm{ep} \frac{\partial \psi_\mrm{p}}{\partial I_3^\mrm{ep}} \mvec{C}_\mrm{p} \right)
		\end{align}

\noindent The state vector of the numerical implementation for the hyperviscoplastic model is:

\begin{equation}
\mvec{z} = \left[ \widetilde{\mvec{S}}^\mrm{T}, \mvec{C}_\mrm{v1}^\mrm{T}, \mvec{C}_\mrm{v2}^\mrm{T}, \mvec{C}_\mrm{p}^\mrm{T} \right]^\mrm{T}
\end{equation}

\noindent To solve the state vector, the Jacobian is of size $4\times4$. The components of the hyperviscoplastic Jacobian are:

\begin{align}
 & \mmat{J}_{11} = \frac{\partial \mvec{r}_1}{\partial \widetilde{\mvec{S}}} = \mmat{I} \label{eq:drds}\\
	&\begin{aligned} \label{eq:j12}
  \mvec{J}_{12} & = \frac{\partial \mvec{r}_1}{\partial \mvec{C}_\mrm{v1}} \\
  & =  \frac{2}{C_1} \left[ \frac{\partial ^2 \psi_\mrm{v1}}{\partial (I_1^\mrm{ev1})^2} \mvec{C}_\mrm{v1}^{-1} \otimes \mvec{C} + \frac{\partial ^2 \psi_\mrm{v1}}{\partial I_1^\mrm{ev1} \partial I_3^\mrm{ev1}} I_3^\mrm{ev1} \left(\mvec{C}_\mrm{v1}^{-1} \otimes \mvec{C}_\mrm{v1}+\mvec{C}^{-1} \otimes \mvec{C} \right) + \right.\\
  & \left. + \left( \frac{\partial ^2 \psi_\mrm{v1}}{\partial (I_3^\mrm{ev1})^2} I_3^\mrm{ev1} + \frac{\partial \psi_\mrm{v1}}{\partial I_3^\mrm{ev1}}\right) I_3^\mrm{ev1} \mvec{C}^{-1} \otimes \mvec{C}_\mrm{v1} \right] \left( \mvec{C}_\mrm{v1}^{-1} \odot \mvec{C}_\mrm{v1}^{-1}\right) + \frac{\partial \psi_\mrm{v1}}{\partial I_1^\mrm{ev1}} \mvec{C}_\mrm{v1}^{-1} \odot \mvec{C}_\mrm{v1}^{-1}
  \end{aligned} \\
   &\begin{aligned}
  \mvec{J}_{13} & = \frac{\partial \mvec{r}_1}{\partial \mvec{C}_\mrm{v2}} \\
  & =  \frac{2}{C_1} \left[ \frac{\partial ^2 \psi_\mrm{v2}}{\partial (I_1^\mrm{ev2})^2} \mvec{C}_\mrm{v2}^{-1} \otimes \mvec{C} + \frac{\partial ^2 \psi_\mrm{v2}}{\partial I_1^\mrm{ev2} \partial I_3^\mrm{ev2}} I_3^\mrm{ev2} \left(\mvec{C}_\mrm{v2}^{-1} \otimes \mvec{C}_\mrm{v2}+\mvec{C}^{-1} \otimes \mvec{C} \right) + \right.\\
  & \left. + \left( \frac{\partial ^2 \psi_\mrm{v2}}{\partial (I_3^\mrm{ev2})^2} I_3^\mrm{ev2} + \frac{\partial \psi_\mrm{v2}}{\partial I_3^\mrm{ev2}}\right) I_3^\mrm{ev2} \mvec{C}^{-1} \otimes \mvec{C}_\mrm{v2} \right] \left( \mvec{C}_\mrm{v2}^{-1} \odot \mvec{C}_\mrm{v2}^{-1}\right) + \frac{\partial \psi_\mrm{v2}}{\partial I_1^\mrm{ev2}} \mvec{C}_\mrm{v2}^{-1} \odot \mvec{C}_\mrm{v2}^{-1}
  \end{aligned}\\
  &\begin{aligned} \label{eq:j14}
  \mvec{J}_{14} & = \frac{\partial \mvec{r}_1}{\partial \mvec{C}_\mrm{p}} \\
  & =  \frac{2}{C_1} \left[ \frac{\partial ^2 \psi_\mrm{p}}{\partial (I_1^\mrm{ep})^2} \mvec{C}_\mrm{p}^{-1} \otimes \mvec{C} + \frac{\partial ^2 \psi_\mrm{p}}{\partial I_1^\mrm{ep} \partial I_3^\mrm{ep}} I_3^\mrm{ep} \left(\mvec{C}_\mrm{p}^{-1} \otimes \mvec{C}_\mrm{p}+\mvec{C}^{-1} \otimes \mvec{C} \right) + \right.\\
  & \left. + \left( \frac{\partial ^2 \psi_\mrm{p}}{\partial (I_3^\mrm{ep})^2} I_3^\mrm{ep} + \frac{\partial \psi_\mrm{p}}{\partial I_3^\mrm{ep}}\right) I_3^\mrm{ep} \mvec{C}^{-1} \otimes \mvec{C}_\mrm{p} \right] \left( \mvec{C}_\mrm{p}^{-1} \odot \mvec{C}_\mrm{p}^{-1}\right) + \frac{\partial \psi_\mrm{p}}{\partial I_1^\mrm{ep}} \mvec{C}_\mrm{p}^{-1} \odot \mvec{C}_\mrm{p}^{-1}
  \end{aligned}\\
  & \mvec{J}_{21} = \mvec{J}_{23} = \mvec{J}_{24} = \mvec{J}_{31} = \mvec{J}_{32} = \mvec{J}_{34} = \mvec{J}_{41} = \mvec{J}_{42} = \mvec{J}_{43} = \mmat{0} \displaybreak[0]\\
  & \begin{aligned} \label{eq:j22}
  \mvec{J}_{22} & = \frac{\partial \mvec{r}_2}{\partial \mvec{C}_\mrm{v1}}\\
   & = \left( \frac{1}{\Delta t} -\frac{4}{\eta_\mrm{v1}} I_3^\mrm{ev1} \frac{\partial \psi_\mrm{v1}}{\partial I_3^\mrm{ev1}} \right)\, \mmat{I} + \frac{4}{\eta_\mrm{v1}} \left[ \frac{\partial ^2 \psi_\mrm{v1}}{\partial (I_1^\mrm{ev1})^2} \mvec{C}\otimes \mvec{C}+ I_3^\mrm{ev1} \frac{\partial ^2 \psi_\mrm{v1}}{\partial I_1^\mrm{ev1} \partial I_3^\mrm{ev1}}  \left( \mvec{C} \otimes \mvec{C}_\mrm{v1} + \mvec{C}_\mrm{v1} \otimes \mvec{C} \right) + \right.\\
   &\left. + \left(\frac{\partial^2 \psi_\mrm{v1}}{\partial (I_3^\mrm{ev1})^2} I_3^\mrm{ev1} + \frac{\partial \psi_\mrm{v1}}{\partial I_3^\mrm{ev1}} \right) I_3^\mrm{ev1} \mvec{C}_\mrm{v1} \otimes \mvec{C}_\mrm{v1} \right]\left( \mvec{C}_\mrm{v1}^{-1} \odot \mvec{C}_\mrm{v1}^{-1}\right)
  \end{aligned}\\
  & \begin{aligned} \label{eq:j33}
  \mvec{J}_{33} & = \frac{\partial \mvec{r}_3}{\partial \mvec{C}_\mrm{v2}}\\
   & = \left( \frac{1}{\Delta t} -\frac{4}{\eta_\mrm{v2}} I_3^\mrm{ev2} \frac{\partial \psi_\mrm{v2}}{\partial I_3^\mrm{ev2}} \right)\, \mmat{I} + \frac{4}{\eta_\mrm{v2}} \left[ \frac{\partial ^2 \psi_\mrm{v2}}{\partial (I_1^\mrm{ev2})^2} \mvec{C}\otimes \mvec{C}+ I_3^\mrm{ev2} \frac{\partial ^2 \psi_\mrm{v2}}{\partial I_1^\mrm{ev2} \partial I_3^\mrm{ev2}}  \left( \mvec{C} \otimes \mvec{C}_\mrm{v2} + \mvec{C}_\mrm{v2} \otimes \mvec{C} \right) + \right.\\
   & \left. \left(\frac{\partial^2 \psi_\mrm{v2}}{\partial (I_3^\mrm{ev2})^2} I_3^\mrm{ev2} + \frac{\partial \psi_\mrm{v2}}{\partial I_3^\mrm{ev2}} \right) I_3^\mrm{ev2} \mvec{C}_\mrm{v2} \otimes \mvec{C}_\mrm{v2} \right]\left( \mvec{C}_\mrm{v2}^{-1} \odot \mvec{C}_\mrm{v2}^{-1}\right)
  \end{aligned} \\
  & \begin{aligned} \label{eq:j44}
  \mvec{J}_{44} & = \frac{\partial \mvec{r}_4}{\partial \mvec{C}_\mrm{p}}\\
   & =  \frac{\mmat{I}}{\Delta t} \left( 1 -2 c_\mrm{p} ||\mvec{C} - \mvec{C}^\mrm{t}|| I_3^\mrm{ep} \frac{\partial \psi_\mrm{p}}{\partial I_3^\mrm{ep}} \right) + \frac{2 c_\mrm{p} ||\mvec{C} - \mvec{C}^\mrm{t}||}{\Delta t} \left[ \frac{\partial ^2 \psi_\mrm{p}}{\partial (I_1^\mrm{ep})^2} \mvec{C}\otimes \mvec{C} + I_3^\mrm{ep} \frac{\partial ^2 \psi_\mrm{p}}{\partial I_1^\mrm{ep} \partial I_3^\mrm{ep}}  \left( \mvec{C} \otimes \mvec{C}_\mrm{p} + \mvec{C}_\mrm{p} \otimes \mvec{C} \right) + \right. \\
   & \left. + \left(\frac{\partial^2 \psi_\mrm{p}}{\partial (I_3^\mrm{ep})^2} I_3^\mrm{ep} + \frac{\partial \psi_\mrm{p}}{\partial I_3^\mrm{ep}} \right) I_3^\mrm{ep} \mvec{C}_\mrm{p} \otimes \mvec{C}_\mrm{p} \right]\left( \mvec{C}_\mrm{p}^{-1} \odot \mvec{C}_\mrm{p}^{-1}\right)
  \end{aligned}
\end{align}

\noindent Components of the RHS matrix for global tangent solution:

\begin{align}
& \begin{aligned} \frac{\partial \mvec{r}_1}{\partial \mvec{E}} & =  2 \frac{\partial \mvec{r}_1}{\partial \mvec{C}}\\
 & = - \frac{4}{C_1}\left( \frac{\partial ^2 \psi_0}{\partial (I_1^\mrm{eq0})^2}\mvec{I} \otimes \mvec {I} + \frac{\partial ^2 \psi_0}{\partial I_1^\mrm{eq0} \partial I_3^\mrm{eq0}} I_3^\mrm{eq0}\left( \mvec{C}^{-1} \otimes \mvec {I} + \mvec{I} \otimes \mvec{C}^{-1} \right) + \left(\frac{\partial ^2 \psi_0}{\partial (I_3^\mrm{eq0})^2} I_3^\mrm{eq0} + \frac{\partial \psi_0}{\partial I_3^\mrm{eqo}} \right) I_3^\mrm{eq0} \mvec{C}^{-1} \otimes \mvec{C}^{-1} -\right.\\
 & \left. -  \frac{\partial \psi_0}{\partial I_3^\mrm{eq0}} I_3^\mrm{eq0} \left(\mvec{C}^{-1} \odot \mvec{C}^{-1}\right) + \frac{\partial ^2 \psi_\mrm{v1}}{\partial (I_1^\mrm{ev1})^2} \mvec{C}_\mrm{v1}^{-1} \otimes \mvec{C}_\mrm{v1}^{-1} + \frac{\partial ^2 \psi_\mrm{v1}}{\partial I_1^\mrm{ev1} \partial I_3^\mrm{ev1}} I_3^\mrm{ev1} \left( \mvec{C}^{-1} \otimes \mvec{C}_\mrm{v1}^{-1}  + \mvec{C}_\mrm{v1}^{-1} \otimes \mvec{C}^{-1} \right) +\right.\\
  & \left.+ \left( \frac{\partial ^2 \psi_\mrm{v1}}{\partial (I_3^\mrm{ev1})^2} I_3^\mrm{ev1}  + \frac{\partial \psi_\mrm{v1}}{\partial I_3^\mrm{ev1}}\right) I_3^\mrm{ev1} \mvec{C}^{-1} \otimes \mvec{C}^{-1} - \frac{\partial \psi_\mrm{v1}}{\partial I_3^\mrm{ev1}} I_3^\mrm{ev1} \mvec{C}^{-1} \odot \mvec{C}^{-1} +\right.\\
  & \left.+ \frac{\partial ^2 \psi_\mrm{v2}}{\partial (I_1^\mrm{ev2})^2} \mvec{C}_\mrm{v2}^{-1} \otimes \mvec{C}_\mrm{v2}^{-1} + \frac{\partial ^2 \psi_\mrm{v2}}{\partial I_1^\mrm{ev2} \partial I_3^\mrm{ev2}} I_3^\mrm{ev2} \left( \mvec{C}^{-1} \otimes \mvec{C}_\mrm{v2}^{-1}  + \mvec{C}_\mrm{v2}^{-1} \otimes \mvec{C}^{-1} \right) + \right.\\
  & \left. + \left( \frac{\partial ^2 \psi_\mrm{v2}}{\partial (I_3^\mrm{ev2})^2} I_3^\mrm{ev2} +
 \frac{\partial \psi_\mrm{v2}}{\partial I_3^\mrm{ev2}}\right) I_3^\mrm{ev2} \mvec{C}^{-1} \otimes \mvec{C}^{-1} - \frac{\partial \psi_\mrm{v2}}{\partial I_3^\mrm{ev2}} I_3^\mrm{ev2} \mvec{C}^{-1} \odot \mvec{C}^{-1} + \right.\\
  & \left. + \frac{\partial ^2 \psi_\mrm{p}}{\partial (I_1^\mrm{ep})^2} \mvec{C}_\mrm{p}^{-1} \otimes \mvec{C}_\mrm{p}^{-1} + \frac{\partial ^2 \psi_\mrm{p}}{\partial I_1^\mrm{ep} \partial I_3^\mrm{ep}} I_3^\mrm{ep} \left( \mvec{C}^{-1} \otimes \mvec{C}_\mrm{p}^{-1}  + \mvec{C}_\mrm{p}^{-1} \otimes \mvec{C}^{-1} \right) + \right.\\
  &\left.+ \left( \frac{\partial ^2 \psi_\mrm{p}}{\partial (I_3^\mrm{ep})^2} I_3^\mrm{ep} +
 \frac{\partial \psi_\mrm{p}}{\partial I_3^\mrm{ep}}\right) I_3^\mrm{ep} \mvec{C}^{-1} \otimes \mvec{C}^{-1} - \frac{\partial \psi_\mrm{p}}{\partial I_3^\mrm{ep}} I_3^\mrm{ep} \mvec{C}^{-1} \odot \mvec{C}^{-1}\right) 
 \end{aligned} \label{eq:pls_dr1de}
 \end{align}

\begin{align}
&\begin{aligned}
	 \frac{\partial \mvec{r}_2}{\partial \mvec{E}} & = 2 \frac{\partial \mvec{r}_2}{\partial\mvec{C}} \\
 & = - \frac{8}{\eta_\mrm{v}}\left[ \frac{\partial ^2 \psi_\mrm{v}}{\partial (I_1^\mrm{ev})^2}  \mvec{C} \otimes \mvec{C}_\mrm{v}^{-1} + I_3^\mrm{ev} \frac{\partial ^2 \psi_\mrm{v}}{\partial I_1^\mrm{ev} \partial I_3^\mrm{ev}} \left( \mvec{C} \otimes \mvec{C}^{-1} + \mvec{C}_\mrm{v} \otimes \mvec{C}_\mrm{v}^{-1} \right) + \frac{\partial \psi_\mrm{v}}{\partial I_1^\mrm{ev}} \mmat{I} + \right.\\
 & \left. + I_3^\mrm{ev} \left( \frac{\partial ^2 \psi_\mrm{v}}{\partial (I_3^\mrm{ev})^2} I_3^\mrm{ev} + \frac{\partial \psi_\mrm{v} }{\partial I_3^\mrm{ev}} \right) \mvec{C}_\mrm{v} \otimes \mvec{C}^{-1} \right]\label{eq:dr2de}
	 \end{aligned}\\
 & \begin{aligned}
\frac{\partial \mvec{r}_3}{\partial \mvec{E}} & = 2 \frac{\partial \mvec{r}_3}{\partial\mvec{C}} \\
 & = - \frac{8}{\eta_\mrm{v2}}\left[ \frac{\partial ^2 \psi_\mrm{v2}}{\partial (I_1^\mrm{ev2})^2}  \mvec{C} \otimes \mvec{C}_\mrm{v2}^{-1} + I_3^\mrm{ev2} \frac{\partial ^2 \psi_\mrm{v2}}{\partial I_1^\mrm{ev2} \partial I_3^\mrm{ev2}} \left( \mvec{C} \otimes \mvec{C}^{-1} + \mvec{C}_\mrm{v2} \otimes \mvec{C}_\mrm{v2}^{-1} \right) + \right. \\
  & \left. + \frac{\partial \psi_\mrm{v2}}{\partial I_1^\mrm{ev2}} \mmat{I} + 
I_3^\mrm{ev2} \left( \frac{\partial ^2 \psi_\mrm{v2}}{\partial (I_3^\mrm{ev2})^2} I_3^\mrm{ev2} + \frac{\partial \psi_\mrm{v2} }{\partial I_3^\mrm{ev2}} \right) \mvec{C}_\mrm{v2} \otimes \mvec{C}^{-1} \right]
\end{aligned} \label{eq:dr3de}\\
& \begin{aligned}
\frac{\partial \mvec{r}_4}{\partial \mvec{E}} & = 2 \frac{\partial \mvec{r}_4}{\partial\mvec{C}} \\
 & = -\frac{4 c_\mrm{p}}{\Delta t ||\mvec{C} - \mvec{C}^\mrm{t}||}\left[ \frac{\partial \psi_\mrm{p}}{\partial I_1^\mrm{ep}} \mvec{C} + I_3^\mrm{ep} \frac{\partial \psi_\mrm{p}}{\partial I_3^\mrm{ep}} \mvec{C}_\mrm{p} \right] \otimes \left( \mvec{C}-\mvec{C}^\mrm{t} \right) - \\
 & - 4 c_\mrm{p} \frac{||\mvec{C} - \mvec{C}^\mrm{t}||}{\Delta t} \left[ \frac{\partial ^2 \psi_\mrm{p}}{\partial (I_1^\mrm{ep})^2}  \mvec{C} \otimes \mvec{C}_\mrm{p}^{-1} + I_3^\mrm{ep} \frac{\partial ^2 \psi_\mrm{p}}{\partial I_1^\mrm{ep} \partial I_3^\mrm{ep}} \left( \mvec{C} \otimes \mvec{C}^{-1} + \mvec{C}_\mrm{p} \otimes \mvec{C}_\mrm{p}^{-1} \right) + \right. \\
  & \left. + \frac{\partial \psi_\mrm{p}}{\partial I_1^\mrm{ep}} \mmat{I} + I_3^\mrm{ep} \left( \frac{\partial ^2 \psi_\mrm{p}}{\partial (I_3^\mrm{ep})^2} I_3^\mrm{ep} + \frac{\partial \psi_\mrm{p} }{\partial I_3^\mrm{ep}} \right) \mvec{C}_\mrm{p} \otimes \mvec{C}^{-1} \right]
\end{aligned} \label{eq:pls_dr4de}
		\end{align}

\noindent For the numerical implementation of the hyperviscoplastic model with modified plastic flow rule, three terms need to be updated in the above formulations:

\noindent Residual vector:
	\begin{equation}
	\begin{aligned}
	 \mvec{r}_4 & = \frac{\mvec{C}_\mrm{p} - \mvec{C}_\mrm{p}^\mrm{t}}{\Delta t} - 2 c_p \frac{||\mvec{C} - \mvec{C}^\mrm{t}||}{\Delta t} \left( \frac{\partial \psi_\mrm{p}}{\partial I_1^\mrm{ep}} \mvec{C} + I_3^\mrm{ep} \frac{\partial \psi_\mrm{p}}{\partial I_3^\mrm{ep}} \mvec{C}_\mrm{p} \right) H\left[ \mrm{sign} \left[ \left( \frac{\partial \psi_\mrm{p}}{\partial I_1^\mrm{ep}} \mvec{C} + I_3^\mrm{ep} \frac{\partial \psi_\mrm{p}}{\partial I_3^\mrm{ep}} \mvec{C}_\mrm{p} \right) \left(\mvec{C} - \mvec{C}^\mrm{t} \right) \right] \right]
	 \end{aligned}
	 \end{equation}
\noindent Jacobian:
	 \begin{equation}
	 \begin{aligned} 
  \mvec{J}_{44} & = \frac{\partial \mvec{r}_4}{\partial \mvec{C}_\mrm{p}}\\
   & =  \frac{\mmat{I}}{\Delta t} \left( 1 -2 c_\mrm{p} ||\mvec{C} - \mvec{C}^\mrm{t}|| I_3^\mrm{ep} \frac{\partial \psi_\mrm{p}}{\partial I_3^\mrm{ep}} \right) + \frac{2 c_\mrm{p} ||\mvec{C} - \mvec{C}^\mrm{t}||}{\Delta t} \left[ \frac{\partial ^2 \psi_\mrm{p}}{\partial (I_1^\mrm{ep})^2} \mvec{C}\otimes \mvec{C}+ I_3^\mrm{ep} \frac{\partial ^2 \psi_\mrm{p}}{\partial I_1^\mrm{ep} \partial I_3^\mrm{ep}}  \left( \mvec{C} \otimes \mvec{C}_\mrm{p} + \mvec{C}_\mrm{p} \otimes \mvec{C} \right) + \right.\\
   & \left. + \left(\frac{\partial^2 \psi_\mrm{p}}{\partial (I_3^\mrm{ep})^2} I_3^\mrm{ep} + \frac{\partial \psi_\mrm{p}}{\partial I_3^\mrm{ep}} \right) I_3^\mrm{ep} \mvec{C}_\mrm{p} \otimes \mvec{C}_\mrm{p} \right]\left( \mvec{C}_\mrm{p}^{-1} \odot \mvec{C}_\mrm{p}^{-1}\right) H\left[ \mrm{sign} \left[ \left( \frac{\partial \psi_\mrm{p}}{\partial I_1^\mrm{ep}} \mvec{C} + I_3^\mrm{ep} \frac{\partial \psi_\mrm{p}}{\partial I_3^\mrm{ep}} \mvec{C}_\mrm{p} \right) \left(\mvec{C} - \mvec{C}^\mrm{t} \right) \right] \right]
  \end{aligned}
	\end{equation}
\noindent RHS matrix for global tangent solution:
	\begin{equation}
	 \begin{aligned}
\frac{\partial \mvec{r}_4}{\partial \mvec{E}} & = 2 \frac{\partial \mvec{r}_4}{\partial\mvec{C}} \\
 & = -\frac{4 c_\mrm{p}}{\Delta t ||\mvec{C} - \mvec{C}^\mrm{t}||}\left[ \frac{\partial \psi_\mrm{p}}{\partial I_1^\mrm{ep}} \mvec{C} + I_3^\mrm{ep} \frac{\partial \psi_\mrm{p}}{\partial I_3^\mrm{ep}} \mvec{C}_\mrm{p} \right] \otimes \left( \mvec{C}-\mvec{C}^\mrm{t} \right) -\\
 & - 4 c_\mrm{p} \frac{||\mvec{C} - \mvec{C}^\mrm{t}||}{\Delta t} \left[ \frac{\partial ^2 \psi_\mrm{p}}{\partial (I_1^\mrm{ep})^2}  \mvec{C} \otimes \mvec{C}_\mrm{p}^{-1} + I_3^\mrm{ep} \frac{\partial ^2 \psi_\mrm{p}}{\partial I_1^\mrm{ep} \partial I_3^\mrm{ep}} \left( \mvec{C} \otimes \mvec{C}^{-1} + \mvec{C}_\mrm{p} \otimes \mvec{C}_\mrm{p}^{-1} \right) + \right. \\
  & \left. \frac{\partial \psi_\mrm{p}}{\partial I_1^\mrm{ep}} \mmat{I} + I_3^\mrm{ep} \left( \frac{\partial ^2 \psi_\mrm{p}}{\partial (I_3^\mrm{ep})^2} I_3^\mrm{ep} + \frac{\partial \psi_\mrm{p} }{\partial I_3^\mrm{ep}} \right) \mvec{C}_\mrm{p} \otimes \mvec{C}^{-1} \right] H\left[ \mrm{sign} \left[ \left( \frac{\partial \psi_\mrm{p}}{\partial I_1^\mrm{ep}} \mvec{C} + I_3^\mrm{ep} \frac{\partial \psi_\mrm{p}}{\partial I_3^\mrm{ep}} \mvec{C}_\mrm{p} \right) \left(\mvec{C} - \mvec{C}^\mrm{t} \right) \right] \right]
\end{aligned}
	\end{equation}	
	

\bibliographystyle{wileyj} 
\bibliography{refs}

\end{document}